\documentclass[graybox]{svmult}

\usepackage{amsmath,amssymb}
\usepackage{comment}
\usepackage{mathptmx}       
\usepackage{helvet}         
\usepackage{courier}        
\usepackage{type1cm}        
\usepackage{makeidx}         
\usepackage{graphicx}        
\usepackage{multicol}        
\usepackage[bottom]{footmisc}
\usepackage[sort&compress,numbers]{natbib}

\usepackage{doi}
\usepackage{hyperref}

\makeindex             

\renewcommand{\epsilon}{\text{\usefont{OML}{cmr}{m}{n}\symbol{15}}}
\newcommand{\eps}{\varepsilon }

\begin{document}

\title*{Enhancement factors for positron annihilation on valence and core orbitals of noble-gas atoms}
\titlerunning{Enhancement factors for positron annihilation on noble-gas atoms} 
\author{D. G. Green and G. F. Gribakin}
\institute{D. G. Green \at Centre for Theoretical Atomic, Molecular and Optical Physics, Queen's University Belfast, Belfast, Northern Ireland, BT71NN, United Kingdom, \email{d.green@qub.ac.uk}
\and G. F. Gribakin \at Centre for Theoretical Atomic, Molecular and Optical Physics, Queen's University Belfast, Belfast, Northern Ireland, BT71NN, United Kingdom,  \email{g.gribakin@qub.ac.uk}}
\maketitle

\abstract{Annihilation momentum densities and correlation enhancement factors for low-energy positron annihilation on valence and core electrons of noble-gas atoms are calculated using many-body theory. s, p and d-wave positrons of momenta up to the positronium-formation threshold of the atom are considered. The enhancement factors parametrize the effects of short-range electron-positron correlations which increase the annihilation
probability beyond the independent-particle approximation. For all positron partial waves and electron subshells, the enhancement factors are found to be relatively insensitive to the positron momentum. The enhancement factors for the core electron orbitals are also independent of the positron angular momentum. The largest enhancement factor ($\sim 15$) is found for the 5p orbital in Xe, while the values for the core orbitals are typically
$\sim 1.5$.}

\section{Introduction}
Low-energy positrons annihilate in atoms and molecules forming two $\gamma$ rays whose Doppler-broadened spectrum is characteristic of the electron velocity distribution in the states involved, and thus of the electron environment. This makes positrons a unique probe in materials science. For example, vacancies and defects in semiconductors and other industrially important materials can be studied \cite{PhysRevLett.77.2097,PhysRevLett.38.241,PhysRevLett.79.39,PhysRevB.20.3566,PhysRevB.54.2397,RevModPhys.85.1583}. Positron-induced Auger-electron spectroscopy (PAES)  \cite{PhysRevLett.61.2245, Ohdaira1997177,Weiss2007285,nepomuc3,hugreview}, and time-resolved PAES \cite{nepomucref,hugreview} enables studies of surfaces with extremely high sensitivity, including dynamics of catalysis, corrosion, and surface alloying \cite{PhysRevLett.105.207401}. 
The $\gamma$ spectra are also sensitive to the positron momentum at the instant of annihilation. This is exploited in Age-MOmentum Correlation (AMOC) experiments (see, e.g., \cite{amoc,posbeams,Sano:amoc:2015}), in which the $\gamma$ spectra are measured as a function of the positron ``age'' (i.e., time after emission from source). AMOC enables study of positron and positronium cooling (and, more generally, transitions between positron states, e.g., to different trapping states, or via chemical reactions) \cite{amoc,posbeams,Sano:amoc:2015}.

Interpretation of the experiments relies heavily on theoretical input, e.g., in PAES one requires accurate relative annihilation probabilities for core electrons of various atoms \cite{PhysRevB.41.3928}. Such quantities, however, are not easy to calculate, as the annihilation process is strongly affected by short-range electron-positron and long-range positron-atom correlations. These effects significantly enhance the annihilation rates \cite{PhysRevA.51.473, DGG_posnobles} and alter the shape and magnitude of the annihilation $\gamma$ spectra \cite{0953-4075-39-7-008,DGG_hlike, DGG_molgammashort, DGG_molgamma,DGG:2015:core}, compared to independent-particle approximation (IPA) calculations. 

A powerful method that allows for systematic inclusion of the correlations in atomic systems is many-body theory (MBT). MBT enables one to calculate the so-called \emph{enhancement factors} (EF), which quantify
the increase of the electron density at the positron due to the effect of correlations. The EF can be used to correct the IPA annihilation probabilities and $\gamma$-spectra \cite{PhysRevLett.38.241,PhysRevB.41.3928}. The EF are particularly large ($\sim $10) for the valence electrons, but are also significant for the core electrons \cite{PhysRevB.20.883}. 
EF were introduced in early MBT works involving positron annihilation in metals that were based on considering positrons in a homogeneous electron gas \cite{PhysRev.129.1622,PhysRev.155.197}. Subsequently, density functional theories were developed to describe positron states and annihilation in a wider class of condensed-matter systems \cite{PhysRevB.34.3820,RevModPhys.66.841}. These methods usually rely on some input in the form of the correlation energy and EF for the positron in electron gas from MBT \cite{Arponen1979343}. When applied to real, inhomogeneous systems, position-dependent EF can lead to spurious effects in the spectra \cite{PhysRevB.54.2397}, and show deficiencies when benchmarked against more accurate calculations \cite{PhysRevB.65.235103}.

In the context of the positron-atom problem, the MBT calculations provided an accurate and essentially complete picture of low-energy positron interaction with noble-gas atoms \cite{DGG_posnobles}, with excellent agreement between the theoretical results and experimental scattering cross sections and annihilation rates. The MBT work was extended recently \cite{DGG:2015:core} to the $\gamma$-spectra for (thermal) positron annihilation on noble-gas atoms. It provided an accurate description of the measured spectra for Ar, Kr and Xe \cite{PhysRevLett.79.39} and firmly
established the relative contributions of various atomic orbitals to the spectra. The calculations also yielded ``exact'' \emph{ab initio} EF $\bar\gamma_{nl}$ for individual electron orbitals $nl$, and found that they follow a simple scaling with the orbital ionization energy \cite{DGG:2015:core}. 

In this work we provide a more detailed analysis and report EF for annihilation of s-, p- and d-wave positrons with momenta up to the positronium formation threshold. We demonstrate that the EF for a given electron orbital and positron partial wave are insensitive to the positron momentum
(in spite of the strong momentum dependence of the annihilation probability \cite{DGG_posnobles}). Moreover, we show that whilst the EF for the core orbitals are almost independent of the positron angular momenta, those for the valence subshells vary between the positron s, p and d waves.
In addition to their use in correcting IPA calculations of positron annihilation with core electrons in condensed matter, the positron-momentum dependent EF calculated here can be used to determine accurate pick-off annihilation rates for positronium in noble gases \cite{psmbt}.

\section{Theory of positron annihilation in many-electron atoms}
\subsection{Basics}

Consider annihilation of a low-energy ($\eps \sim 1$~eV) positron with momentum ${\bf k}$ in a many-electron system, e.g., an atom. In the dominant process, the positron annihilates with an electron in state $n$ to form two $\gamma$-ray photons of total momentum ${\bf P}$ \cite{QED}. In the centre-of-mass frame, where the total momentum ${\bf P}$ is zero, the two photons are emitted in opposite directions and have equal energies 
$E_{\gamma}=p_{\gamma}c=mc^2+\frac{1}{2}(E_{\rm i}-E_{\rm f})\simeq mc^2\simeq 511\,{\rm keV}$,
where $E_{\rm i}$ and $E_{\rm f}$ denote the energy of the initial and final states of the system (excluding rest mass). 
When ${\bf P}$ is non-zero, however, the two photons no longer propagate in exactly opposite directions
and their energy is Doppler shifted. For example, for the first photon
$E_{\gamma_1}=E_{\gamma}+mcV\cos{\theta}$, where $\theta$ is the angle between the momentum of the photon and the centre-of-mass velocity of the electron-positron pair ${\bf V}={\bf P}/2m$ (assuming that $V\ll c$, and $p_{\gamma_1}=E_{\gamma_1}/c\approx mc$). The Doppler shift of the photon energy from the centre of the line then is
\begin{eqnarray}
\epsilon=E_{\gamma_1}-E_{\gamma}=mc\;V\cos{\theta}=\frac{Pc}{2}\cos{\theta}.
\end{eqnarray}
The typical momenta of electrons bound with energy $\eps_n$ determine the characteristic width of the annihilation spectrum $\epsilon \sim Pc \sim\sqrt{|\eps_n|mc^2}\gg|\eps_n|$. 
Hence the shift $\eps_n/2$ of the line centre $E_\gamma $ from $mc^2=511$\,keV can usually be neglected, even for the core electrons. The $\gamma$ spectrum averaged over the direction of emitted photons (or that of the positron momentum ${\bf k}$) takes the form (see, e.g., \cite{0953-4075-39-7-008})
\begin{eqnarray}\label{eqn:gammaspectra}
w_n(\epsilon)=\frac{1}{c} \int \int _{2|\epsilon|/c}^{\infty}
|A_{n{\bf k}}({\bf P})|^2\frac{PdPd\Omega_{\bf P}}{(2\pi)^3},
\end{eqnarray}
where 
$A_{n{\bf k}}({\bf P})$ is the annihilation amplitude, whose calculation using MBT is described below. The quantity $|A_{n{\bf k}}({\bf P})|^2$
is the annihilation momentum density (AMD)\footnote{Alternatively to the Doppler-shift spectrum, experiments measure the one-dimensional angular correlation of annihilation radiation (1D-ACAR), i.e., the small angle
$\Theta $ small between the direction of one photon and the plane containing
the other. The corresponding distribution can be obtained from $w(\eps )$ 
using $\Theta =2\epsilon /mc^2$. Not also that if the positron wavefunction is constant, then the annihilation momentum density is proportional to the electron momentum density, and the $\gamma$ spectrum becomes similar to the Compton profile \cite{Kaijser197737,DGG_molgammashort,DGG_molgamma}.}.

The annihilation rate $\lambda$ for a positron in a gas of atoms or molecules with number density $n_m$ is usually parametrized by
\begin{eqnarray}
\lambda = \pi r_0^2cn_m Z_{\rm eff},
\end{eqnarray}
where $r_0=e^2/mc^2$ is the classical radius of the electron (in CGS units) and $Z_{\rm eff}$ is the effective number of electrons per target atom or molecule that contribute to annihilation \cite{Fraser,pomeranchuk}. It is found as a sum over electron states $Z_{\rm eff}=\sum_nZ_{{\rm eff},n}$, where
\begin{eqnarray}\label{eq:Zeff}
Z_{{\rm eff},n}=\int {w}_{n}(\epsilon)\,d\epsilon =
\int |A_{n{\bf k}}({\bf P})|^2\frac{d^3{\bf P}}{(2\pi)^3}
\end{eqnarray}
is the partial contribution due to positron annihilation with electron in state $n$, and where it is assumed that the incident positron wavefunction used in the calculation of $A_{n{\bf k}}({\bf P})$ is normalized to a plane wave. 
In general, the parameter $Z_{\rm eff}$ is different from the number of electrons in the target atom $Z$. In particular, positron-atom and electron-positron correlations can make $Z_{\rm eff}\gg Z$ \cite{DGG_posnobles,DGG:2015:core,PhysScripta.46.248,dzuba_mbt_noblegas,0953-4075-38-6-R01}.

\subsection{Many-body theory for the annihilation amplitude}

The incident positron wavefunction is taken in the form of a partial-wave
expansion\footnote{In this and subsequent sections we make wide use of atomic units (a.u.).}
\begin{eqnarray}\label{eq:poswf}
\psi_{{\bf k}}({\bf r})=\frac{4\pi}{{r}}\sqrt{\frac{\pi}{k}}\sum_{\ell m}i^{\ell}e^{i\delta_{\ell}}
Y^\ast_{\ell m}(\hat{\bf k}) Y_{\ell m}(\hat{\bf r})
P_{\eps \ell}(r),
\end{eqnarray} 
where $\delta_{\ell}$ is the scattering phaseshift \cite{quantummechanics}, $Y_{\ell m}$ is the spherical harmonic, and where the radial function with orbital angular momentum $\ell$ is normalized by its asymptotic behaviour $P_{\eps \ell}(r)\simeq (\pi k)^{-1/2}\sin(kr -\pi \ell /2+\delta_{\ell})$. In the simplest approximation the radial wavefunctions are calculated in the static field of the ground-state (Hartree-Fock, HF) atom. This approximation is very inaccurate for the positron-atom problem. It fails to describe the scattering cross sections and grossly underestimates the annihilation rates. Much more accurate positron wavefunctions (Dyson orbitals) are obtained by solving the Dyson equation which includes the nonlocal, energy-dependent positron-atom correlation potential \cite{DGG_posnobles,PhysRevA.70.032720} (Sec.~\ref{subsubsec:Dyson}).

In the lowest-order approximation the annihilation amplitude is given by
\begin{eqnarray}\label{eq:Ank0}
A_{n{\bf k}}({\bf P})=\int e^{-i{\bf P}\cdot {\bf r}}\psi _{\bf k}({\bf r})\varphi _n({\bf r}) d^3{\bf r},
\end{eqnarray}
where $\varphi _n({\bf r})\equiv \varphi _{nlm}({\bf r})=\frac{1}{r}P_{nl}(r)Y_{lm}(\hat{\bf r})$ is the wavefunction of electron in subshell $nl$. Equation (\ref{eq:Ank0}) is equivalent to IPA. After integration over the directions of ${\bf P}$ in the spectrum (\ref{eqn:gammaspectra}), all positron partial waves contribute to the AMD incoherently. This means that the annihilation amplitude can be calculated independently for each $\ell$, replacing $\psi _{\bf k}({\bf r})$ in Eq.~(\ref{eq:Ank0}) by the corresponding positron partial wave orbital $\psi _\eps ({\bf r})$. Omitting the index $\ell $, we denote such amplitude $A_{n\eps }({\bf P})$.

As described below, the main corrections to the zeroth-order amplitude originate from the electron-positron Coulomb interaction which increases the probability of finding the electron and positron at the same point in space.

\subsubsection{The annihilation vertex}

\begin{figure}[t!]
\sidecaption[t]
\includegraphics[width=\textwidth]{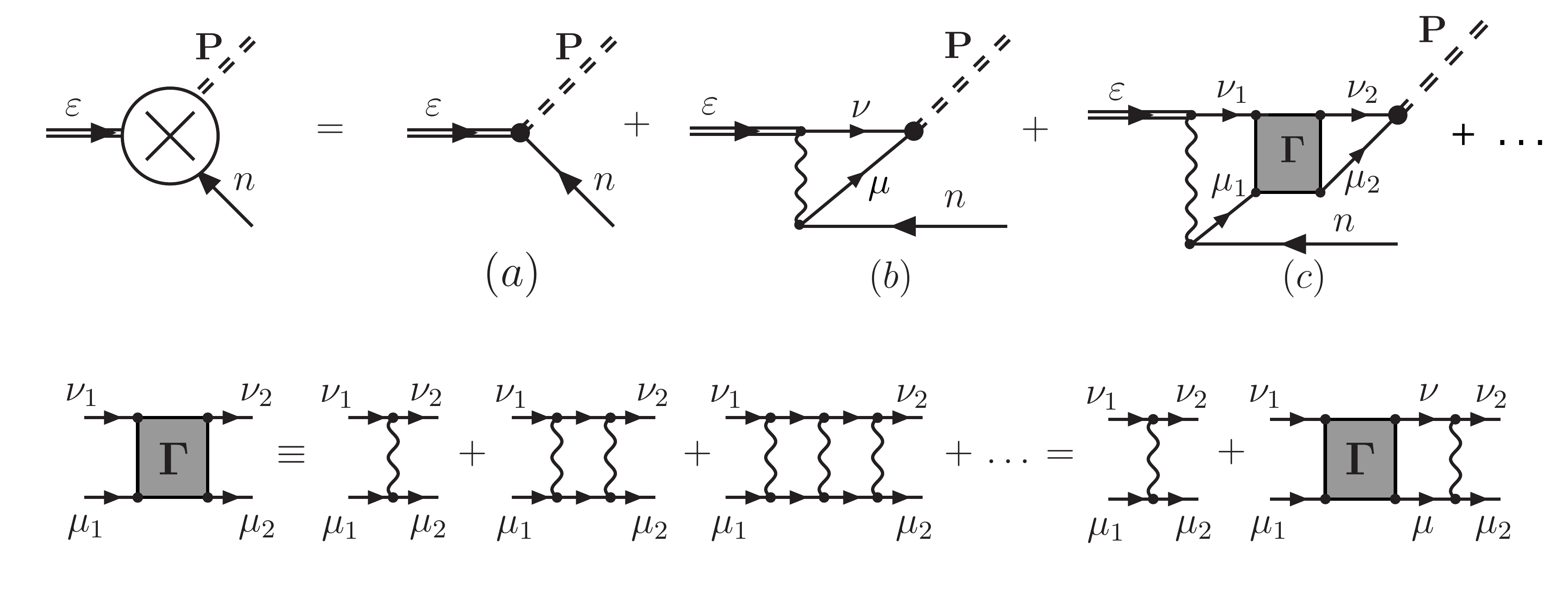}
\caption{Amplitude of positron annihilation with an electron in state $n$:
(a) zeroth-order, (b) first-order, and (c) with virtual-positronium corrections. Double lines labelled $\eps$ represent the incident positron; single lines labelled $\nu$ (${\mu}$) represent positron (excited electron) states; lines labelled $n$ represent holes in the atomic ground state; wavy lines represent the electron-positron Coulomb interaction, and double-dashed lines represent the two $\gamma$-ray photons. The $\Gamma$-block is the sum of the electron-positron ladder diagram series. Summation over all intermediate positron, electron, and hole states is assumed.
\label{fig:anndiags}}       
\end{figure}

Figure \ref{fig:anndiags} shows the amplitude $A_{n\eps}({\bf P})$ in diagrammatic form \cite{0953-4075-39-7-008,DGG:2015:core,DGG_thesis,DGG_hlike}\footnote{It is also possible to develop a diagrammatic expansion for $Z_{\rm eff}$ \cite{PhysScripta.46.248,dzuba_mbt_noblegas,PhysRevA.70.032720,0953-4075-39-7-008,DGG_posnobles} that enables one to calculate the annihilation rate directly, rather than from Eq.~(\ref{eq:Zeff}).}. The total amplitude is depicted on left-hand side of the diagrammatic equation, with the double line ($\eps$) corresponding to incoming positron that annihilates an electron in orbital $n$, producing two $\gamma $-rays (double-dashed line), and the circle with a cross denoting the full annihilation vertex. The main contributions to the amplitude are shown on the right-hand side of the equation. Diagram (a) is the zeroth-order amplitude [IPA, Eq.~(\ref{eq:Ank0})], diagram (b) is the
first-order correction and diagram (c) is the nonperturbative `virtual-positronium' correction. This correction contains the shaded `$\Gamma$-block' which represents the sum of an infinite series of electron-positron ladder diagrams shown in the lower part of Fig.~\ref{fig:anndiags}.

The ladder diagrams represented by the $\Gamma$-block are important because the electron-positron Coulomb attraction supports bound states of the positronium (Ps) atom. To form Ps the energy of the incident positron needs to be greater than the Ps-formation threshold $E_{\rm Ps}=I-6.8~\text{eV}$, where $I$ is the ionization potential of the atom. However, even at lower energies where the Ps can only be formed virtually, this process gives a noticeable contribution. In practice, the $\Gamma$-block is found from the linear equation $\Gamma = V + V\chi\Gamma$, shown diagrammatically in the lower part of Fig.~\ref{fig:anndiags}, where $V$ is the electron-positron Coulomb interaction and $\chi$ is the propagator of the intermediate electron-positron state. Discretizing the electron and positron continua by confining the system in a spherical cavity reduces this to a linear matrix equation, which is easily solved numerically (see \cite{PhysRevA.70.032720,DGG_hlike,DGG_posnobles,DGG_corelong} for further details).  

The total amplitude takes the form
\begin{eqnarray}\label{eqn:annampgeneral}
A_{n\eps}({\bf P})=\int e^{-i{\bf P}\cdot{\bf r}}\left\{ \psi_{\eps}({\bf r})\varphi_n({\bf r})\right.
&+ \left.\tilde\Delta _\eps ({\bf r};{\bf r}_1,{\bf r}_2)\psi_{\eps}({\bf r}_1)\varphi_n({\bf r}_2) d^3{\bf r}_1d^3{\bf r}_2\right\} d^3{\bf r}.
\end{eqnarray}
Here, the first term, corresponding to Fig.~\ref{fig:anndiags} (a), is simply the Fourier transform of the product of electron and positron wavefunctions, taken at the same point. The second term, involving the non-local annihilation kernel $\tilde\Delta _\eps $ (of non-trivial form), describes the vertex corrections. Note that $A_{n\eps}(\bf P)$ is the Fourier transform of the correlated pair wavefunction (the term in the braces). Refs. \cite{DGG_corelong,DGG_thesis} present the partial-wave analysis and corresponding working analytic expressions for the matrix elements involving the vertex corrections.

\subsubsection{Dyson equation for the positron wavefunction}\label{subsubsec:Dyson}

As mentioned above, accurate annihilation rates and $\gamma $-spectra can be obtained only by taking into account the positron-atom correlation potential. This potential is decribed by another class of diagrams 
that ``dress'' the positron wavefunction. 
The corresponding positron \emph{quasiparticle} wavefunction (or Dyson orbital, double line in Fig.~\ref{fig:anndiags}) is calculated from the Dyson equation (see, e.g., \cite{abrikosov,fetterwalecka,mbtexposed})
\begin{eqnarray}\label{eqn:dyson}
\left(\hat{H}_0+\hat{\Sigma}_{\eps}\right)\psi_{\eps}({\bf r})=\eps\psi_{\eps}({\bf r}).
\end{eqnarray}
Here $\hat{H}_0$ is the Hamiltonian of the positron in the static field of the $N$-electron atom in the ground state (described at the HF level). The nonlocal, energy-dependent correlation potential $\hat{\Sigma}_{\eps}$ is equal to the self-energy of the positron Green's function \cite{PhysRevLett.3.96}, and acts as an integral operator $\hat{\Sigma}_{\eps}\psi_{\eps}({\bf r)}=\int \Sigma_{\eps}({\bf r},{\bf r}')\psi_{\eps}({\bf r'})d^3{\bf r}'$.

\begin{figure}[!t]
\centering
\includegraphics[width=0.9\textwidth]{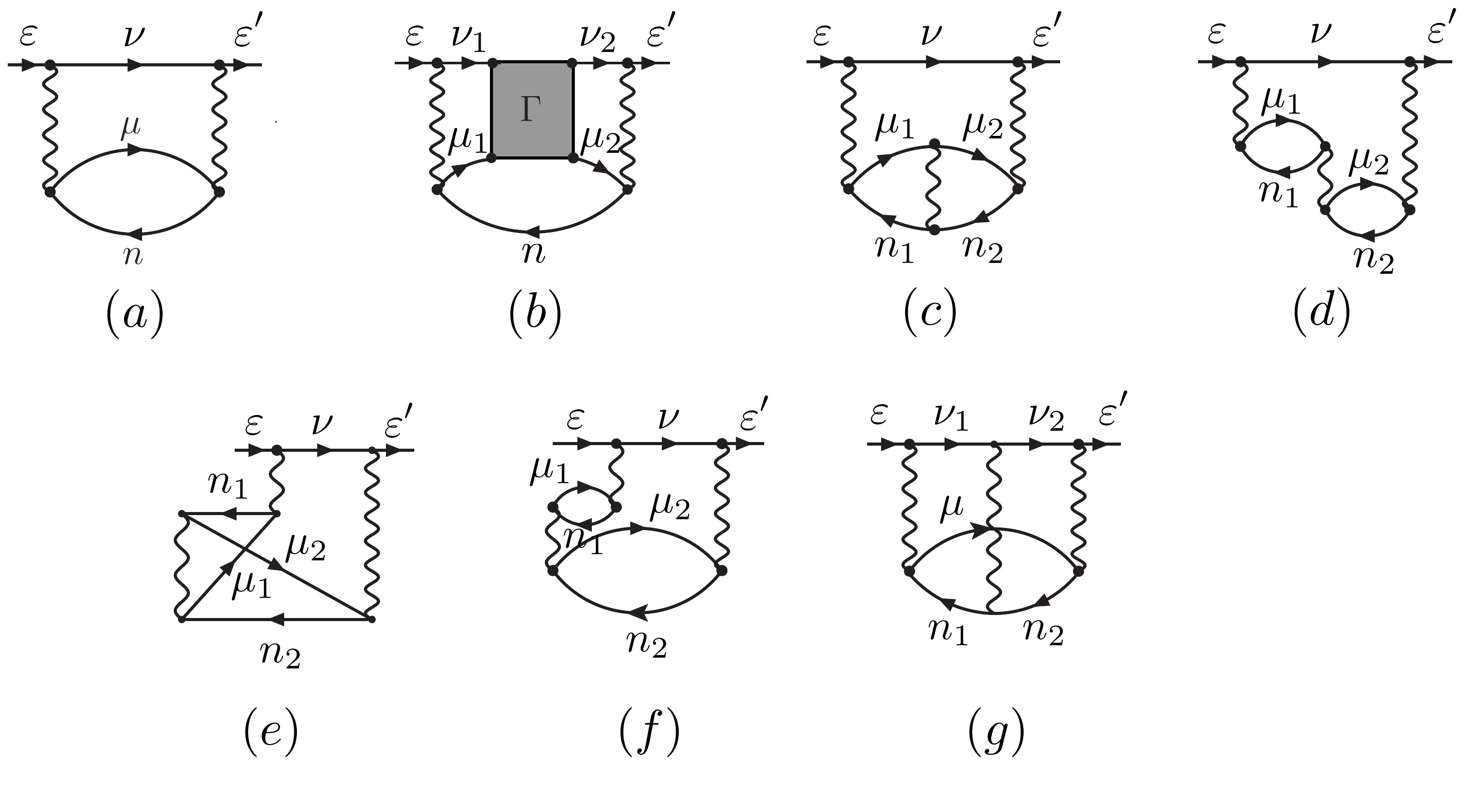}
\caption{
Main contributions to the positron self-energy matrix $\langle \eps '|\hat \Sigma _E|\eps \rangle $. The lowest, second-order diagram (a) describes the effect of polarization; diagram (b) accounts for virtual Ps formation represented by the $\Gamma$-block. Diagrams (c)--(g) represent leading third-order corrections not included in (b). Top lines in the diagrams describe the positron. Other lines with the arrows to the right are excited electron states, and to the left, holes, i.e., electron states occupied in the target ground state. Wavy lines represent Coulomb interactions. Summation over all intermediate states is assumed.
\label{fig:selfenergy}}
\end{figure}

The main contributions to $\hat \Sigma _\eps $ are shown in Fig.~\ref{fig:selfenergy}. At large positron-atom distances the correlation potential reduces to the local polarization potential $\Sigma _\eps ({\bf r},{\bf r}')\simeq -{\alpha_d}\delta({\bf r-r}')/{2r^4}$, where $\alpha_d$ is the dipole polarizability of the atom. If only diagram Fig.~\ref{fig:selfenergy}~(a) is included, the polarizability is given by the HF approximation
\begin{eqnarray}
\alpha_d=\frac{2}{3}\sum_{n,\mu}\frac{|\langle \mu|{\bf r}|n\rangle|^2}{\eps_{\mu}-\eps_n}.
\end{eqnarray}
Diagrams Fig.~\ref{fig:selfenergy} (c), (d), (e) and (f) are third-order corrections to the polarization diagram (a) of the type described by the random-phase approximation with exchange \cite{amusia_casestudiesinatphys1975}. Including these gives asymptotic behaviour of $\Sigma _\eps ({\bf r},{\bf r}')$ with a more accurate value of $\alpha _d$. The digram Fig.~\ref{fig:selfenergy} (b) describes the virtual Ps-formation contribution. Adding it to diagram Fig.~\ref{fig:selfenergy}~(a) nearly doubles the strength of the correlation potential in heavier noble-gas atoms (Ar, Kr and Xe). The diagram Fig. \ref{fig:selfenergy} (d) describes the positron-hole repulsion. Including the diagrams of
Fig.~\ref{fig:selfenergy} in the positron-atom correlation potential provides accurate scattering phaseshifts and cross sections for all noble-gas atoms \cite{DGG_posnobles}.

The positron self-energy diagrams and the annihilation amplitude contain sums over the intermediate excited electron and positron states. In practice we calculate them numerically using sets of electron and positron basis states constructed using 40 B-splines of order 6, in a spherical box of radius 30~a.u. We use an expotential knot sequence for the B-splines, which provides for an efficient spanning of the electron and positron continua in the sums over intermediate states \cite{PhysRevA.70.032720}. The maximum angular momentum of the intermediate states is $l_{\rm max}$=15, and we extrapolate to $l_{\rm max}\to\infty$ as in \cite{PhysRevA.70.032720,DGG_hlike,DGG_posnobles,DGG_corelong}.

\section{Annihilation momentum densities for valence and core electron orbitals in noble gases}

\begin{figure}[h!!]
\includegraphics[width=0.45\textwidth]{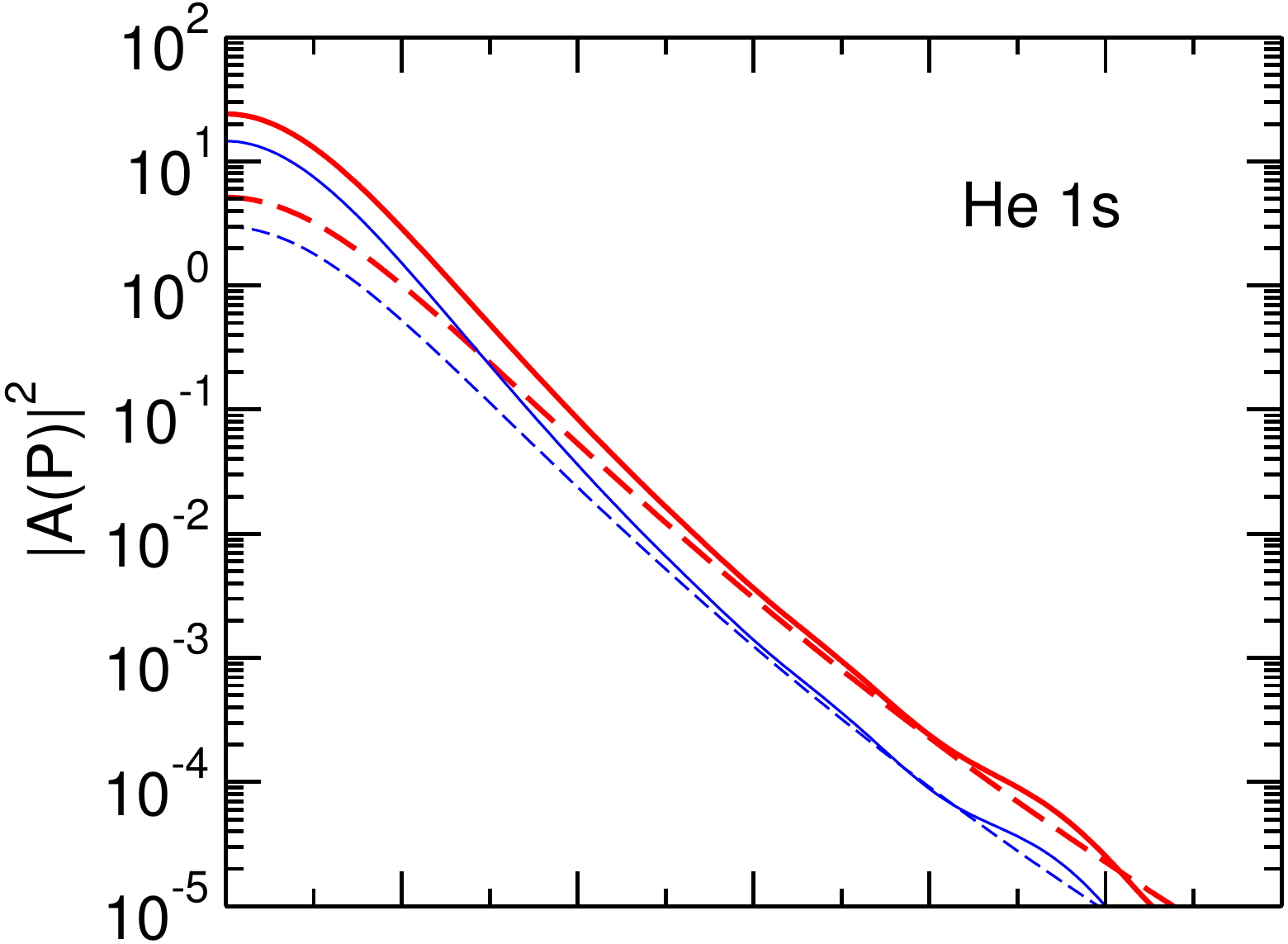}~~~~
\includegraphics[width=0.45\textwidth]{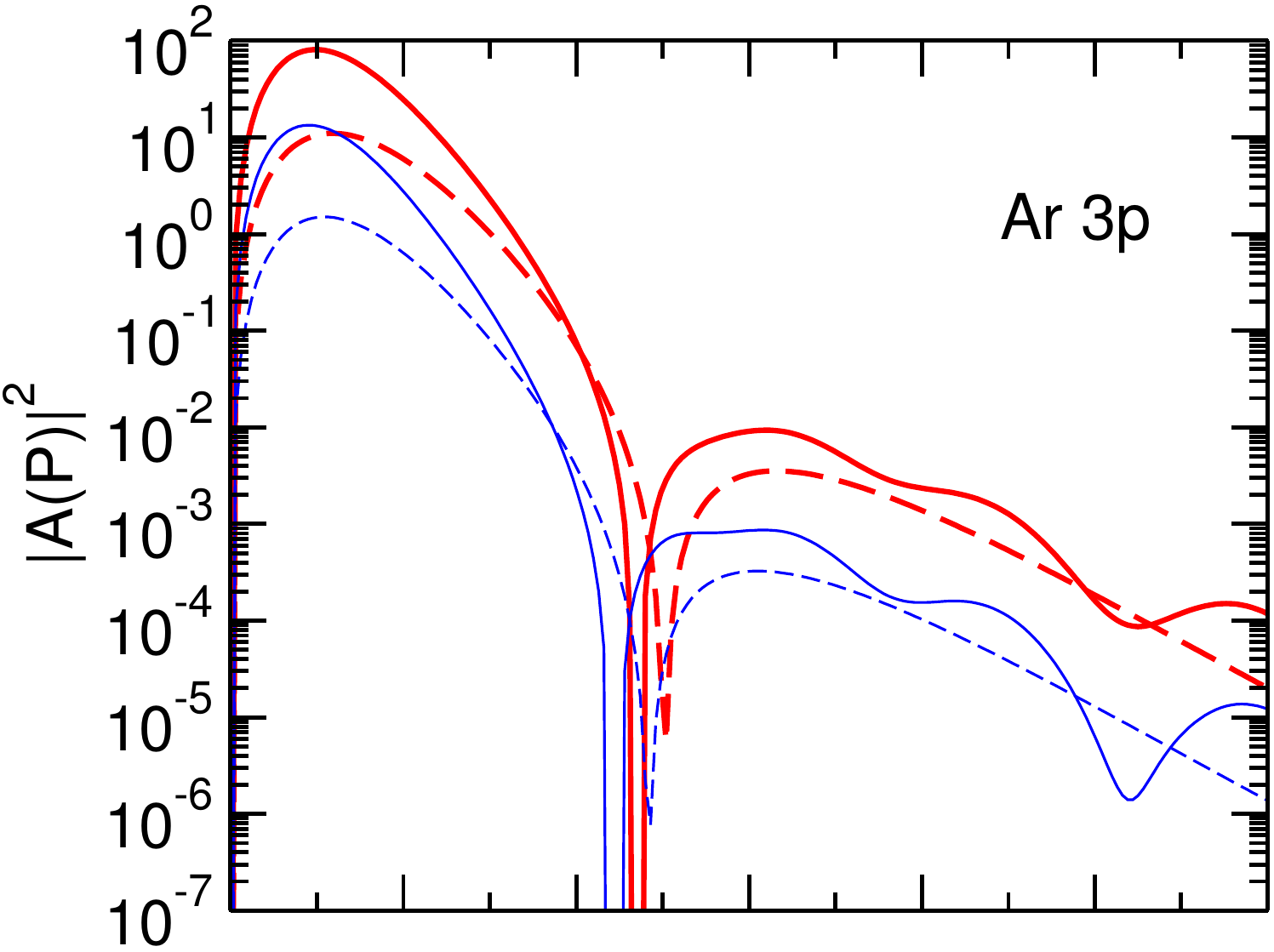}\\
\includegraphics[width=0.45\textwidth]{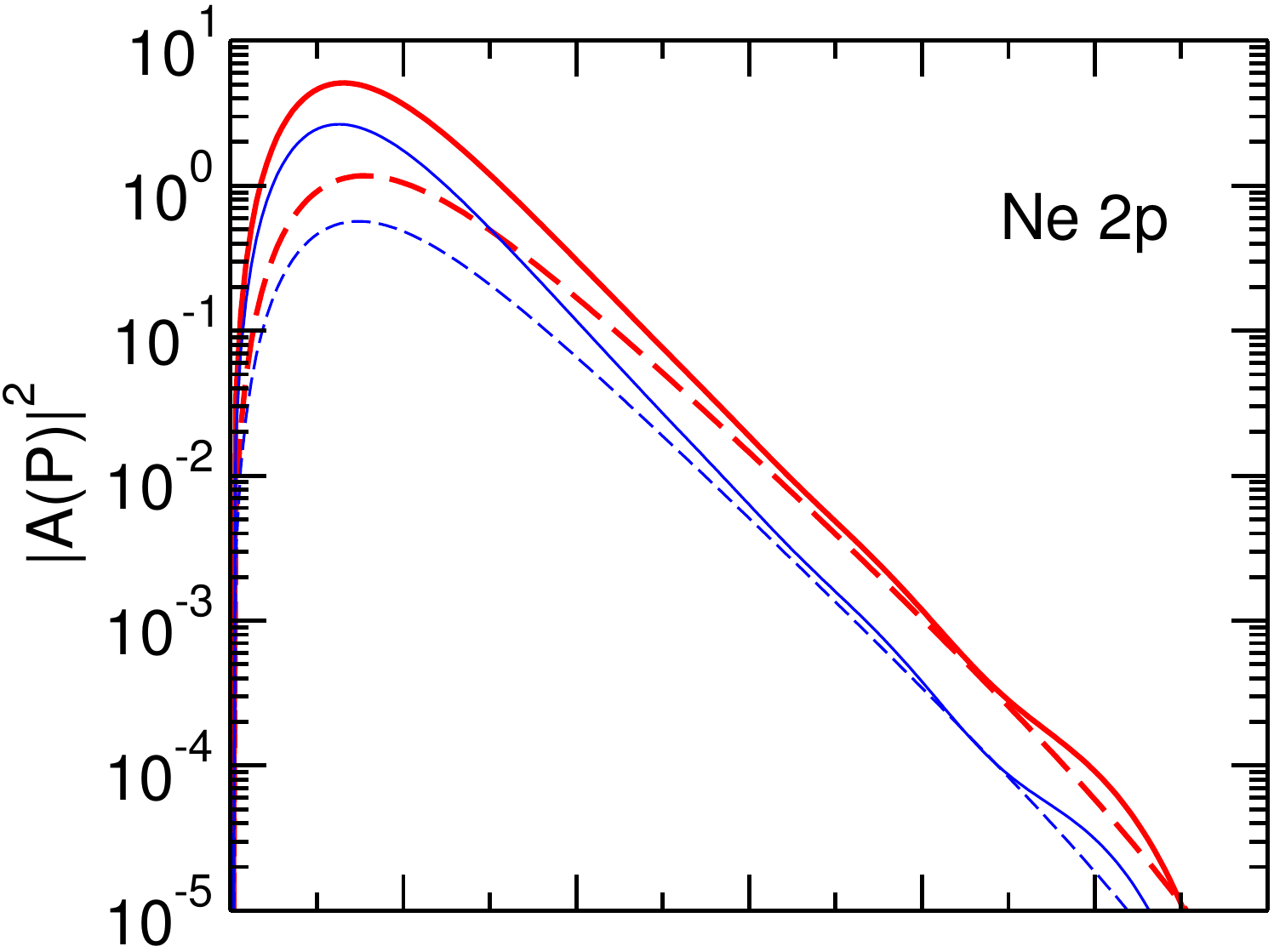}~~~~
\includegraphics[width=0.45\textwidth]{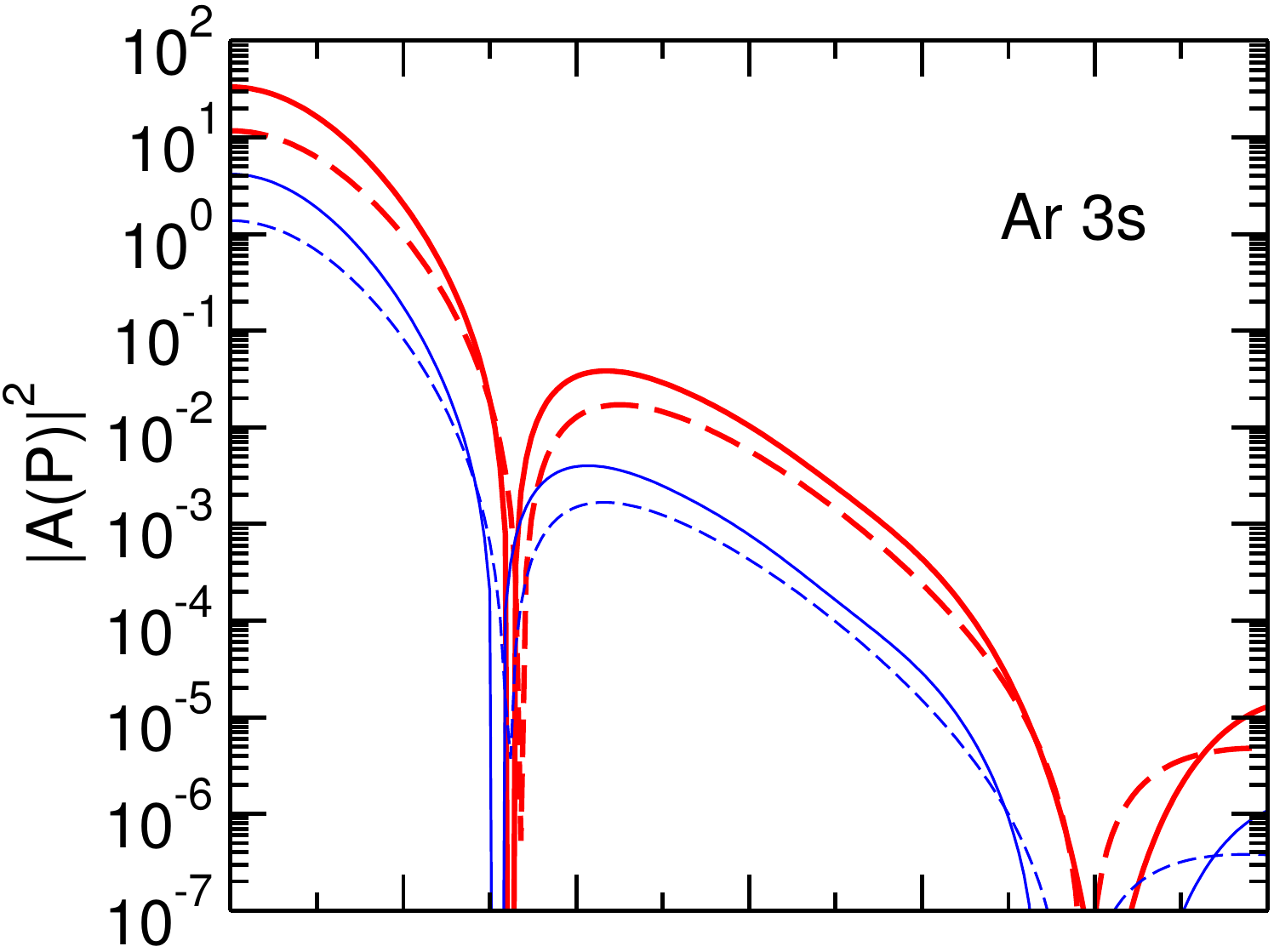}\\%
\includegraphics[width=0.45\textwidth]{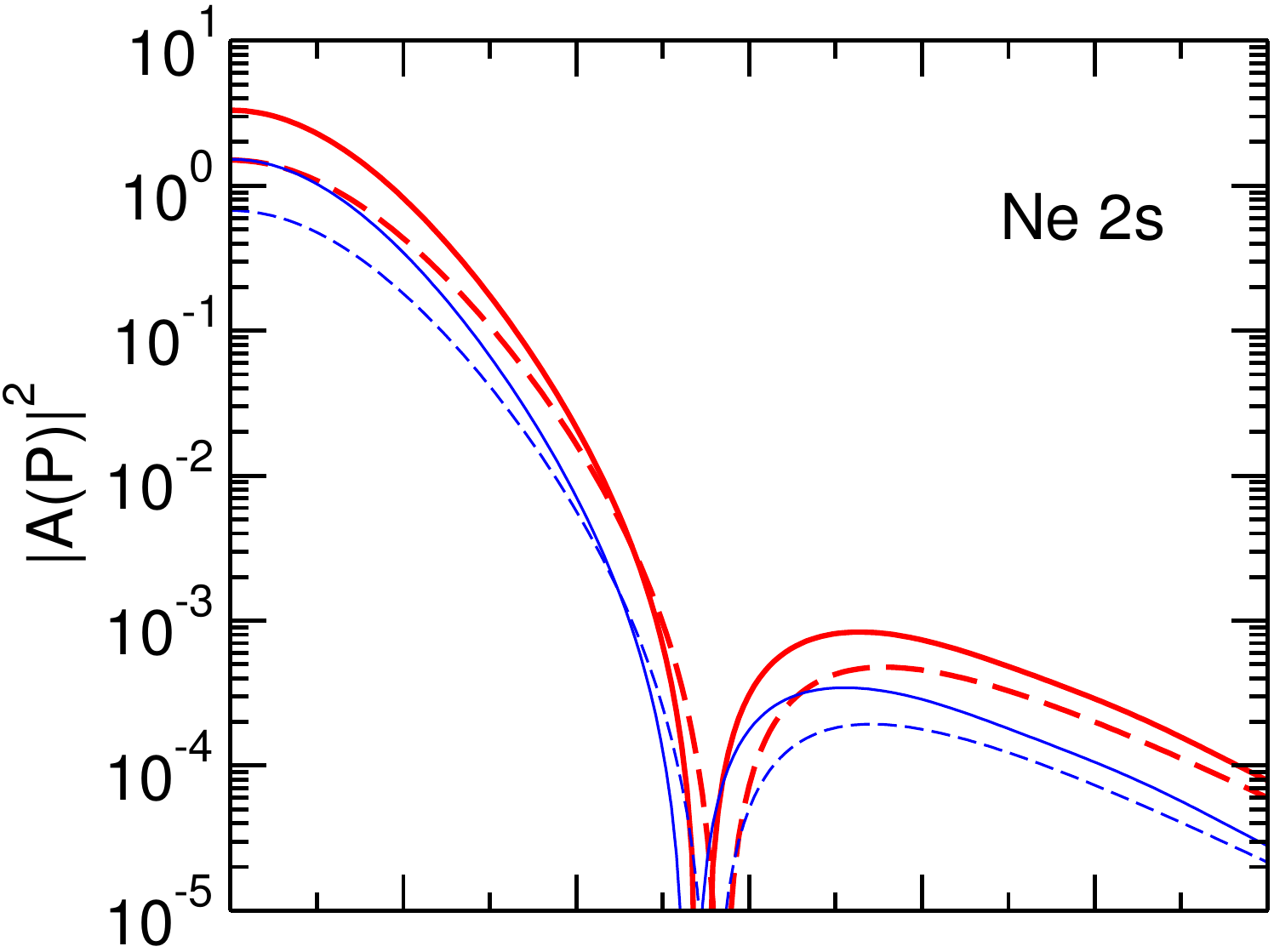}~~~~
\includegraphics[width=0.45\textwidth]{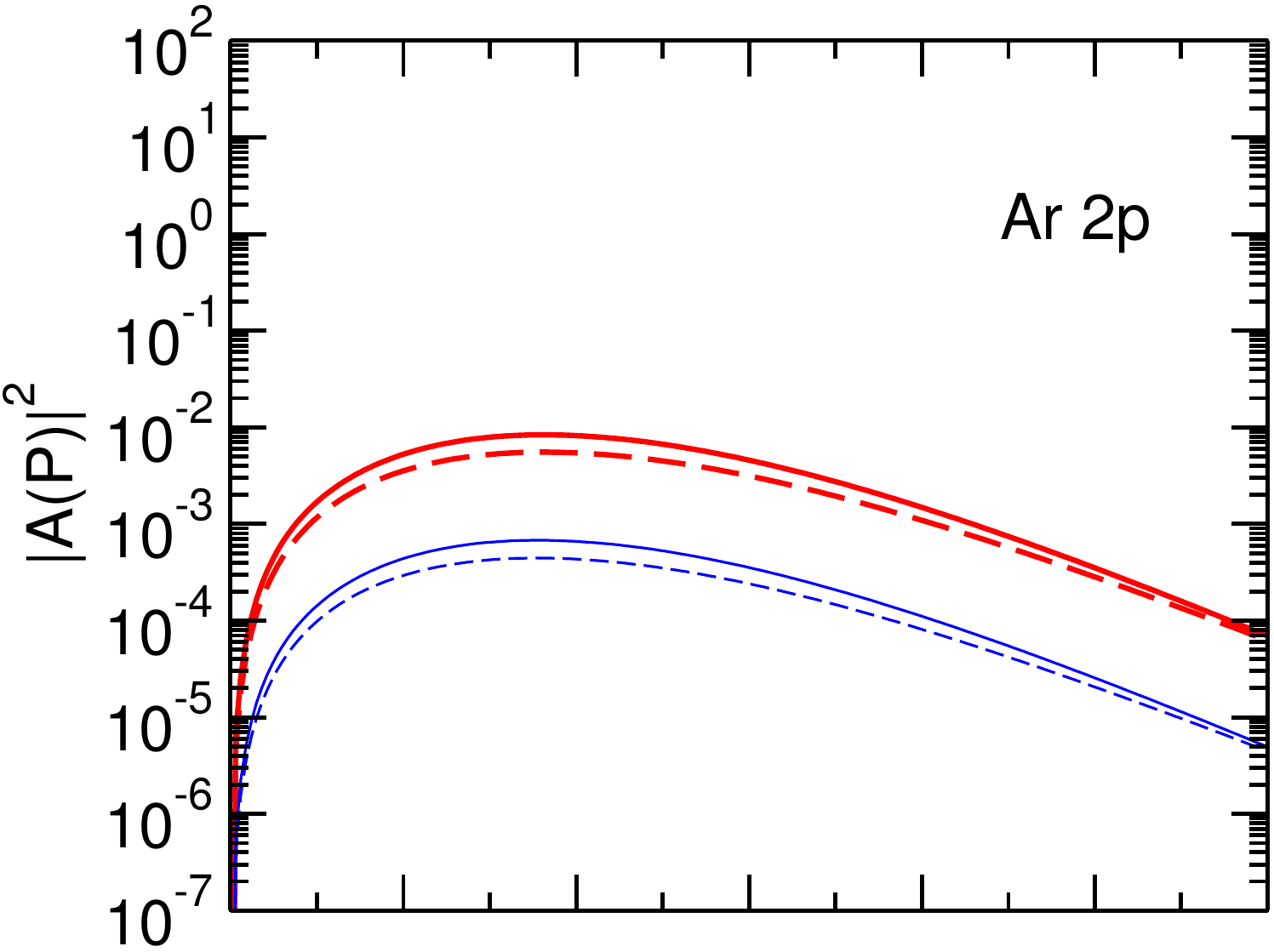}\\
\includegraphics[width=0.45\textwidth]{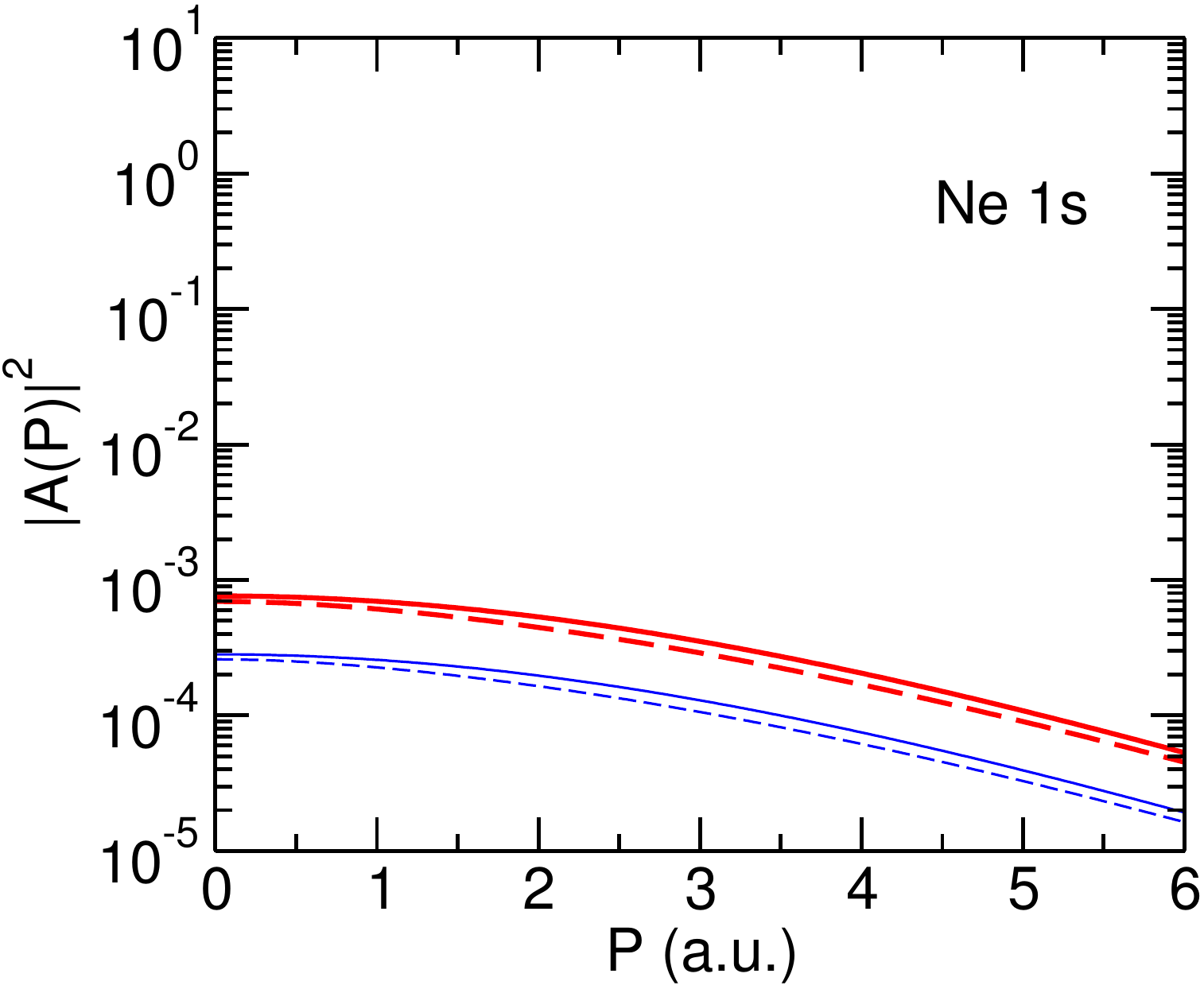}~~~~
\includegraphics[width=0.45\textwidth]{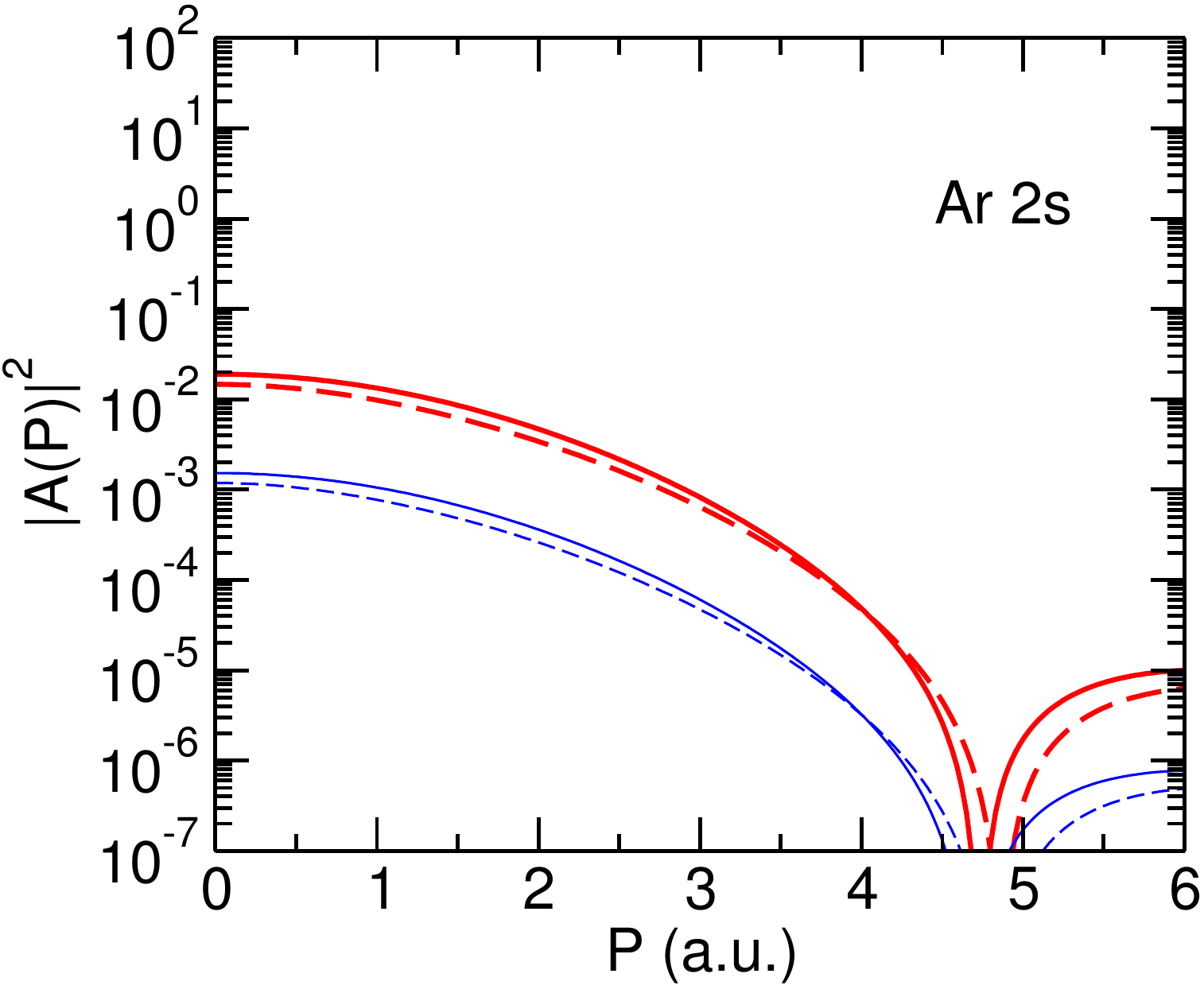}%
\caption{Annihilation momentum density as a function of the total $2\gamma$ momentum $P$, for s-wave positron annihilation in He, Ne and Ar, calculated in different approximations for the annihilation vertex: zeroth-order vertex (dashed lines); full vertex ($0+1+\Gamma$) (solid lines), for HF (thin blue lines) and Dyson (thick red lines) positron of momentum $k=0.04$ a.u. (For a given approximation for the vertex, the lines for the calculation with Dyson positron lie above those for HF positron). }
\label{fig:he}       
\end{figure}

\begin{figure}[p!]
\includegraphics[width=0.45\textwidth]{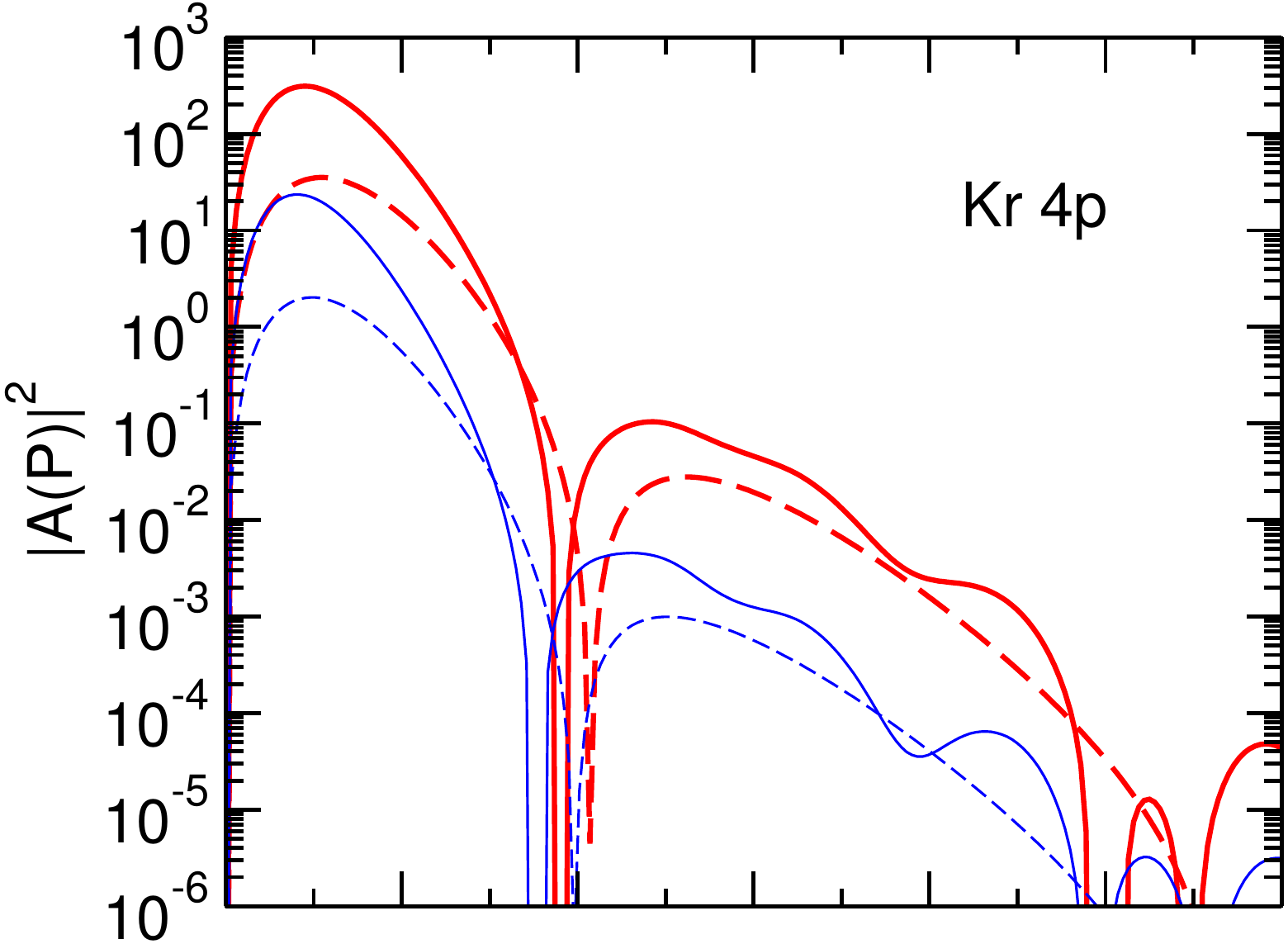}~~~~
\includegraphics[width=0.45\textwidth]{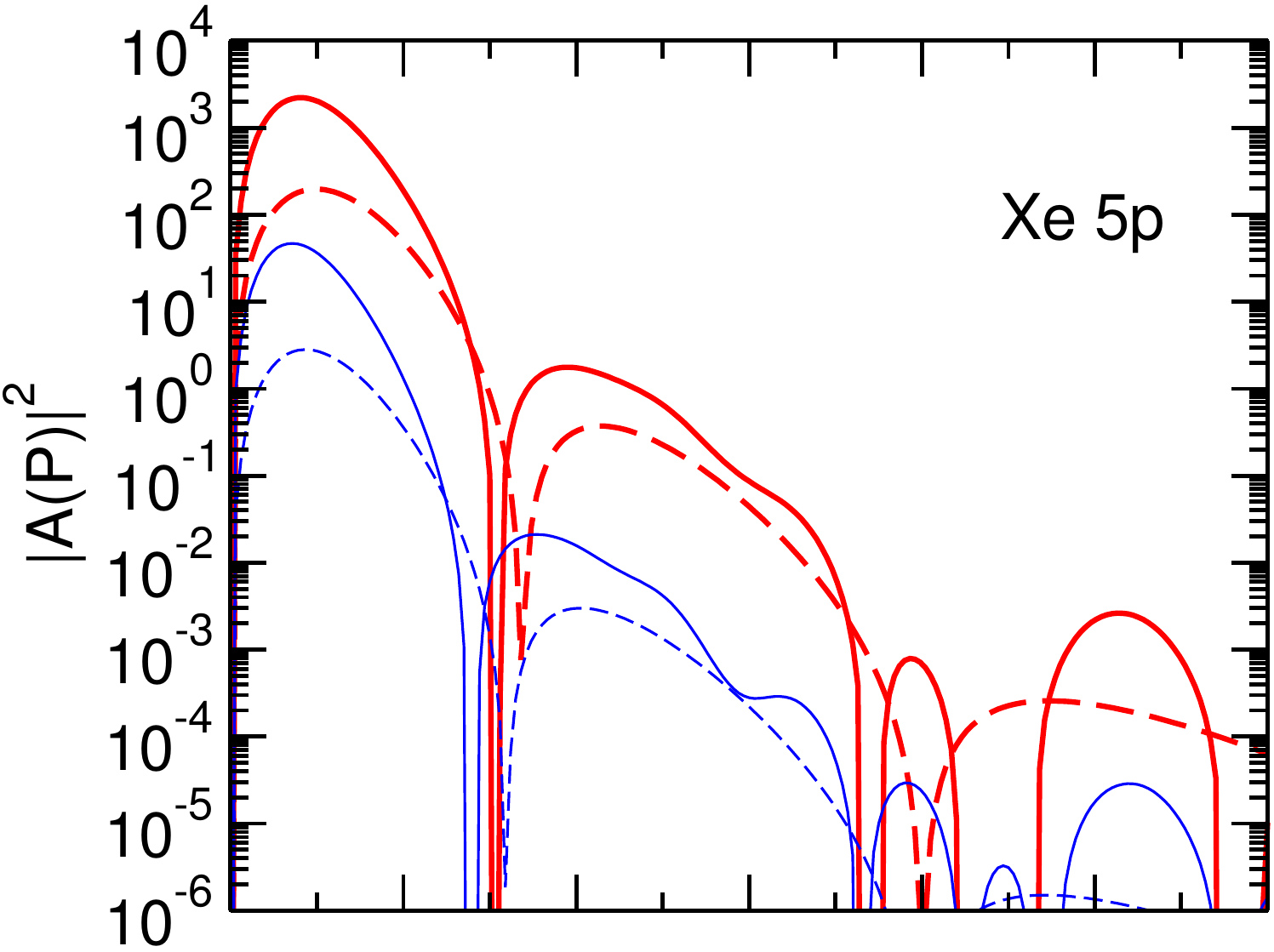}\\
\includegraphics[width=0.45\textwidth]{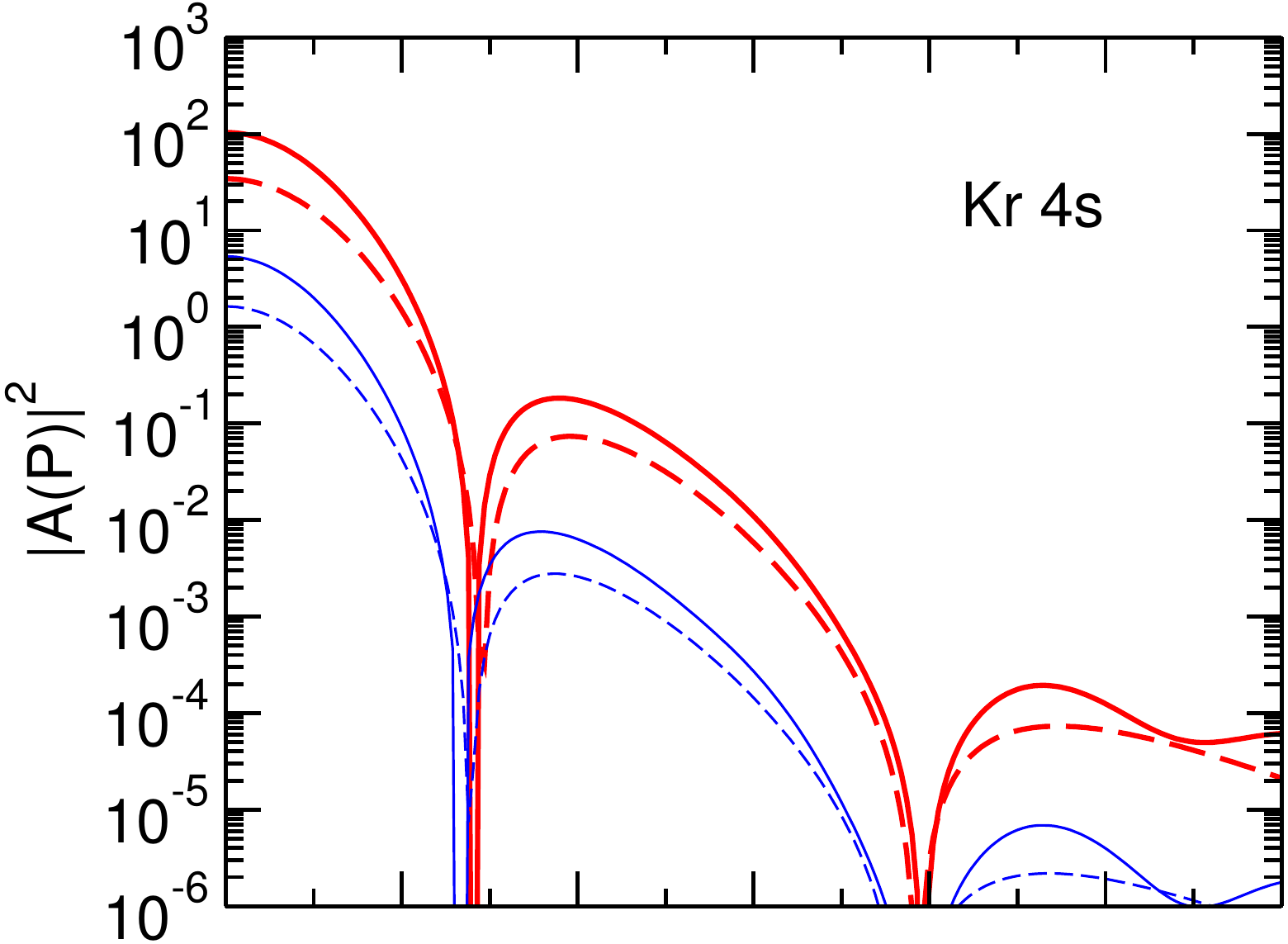}~~~~
\includegraphics[width=0.45\textwidth]{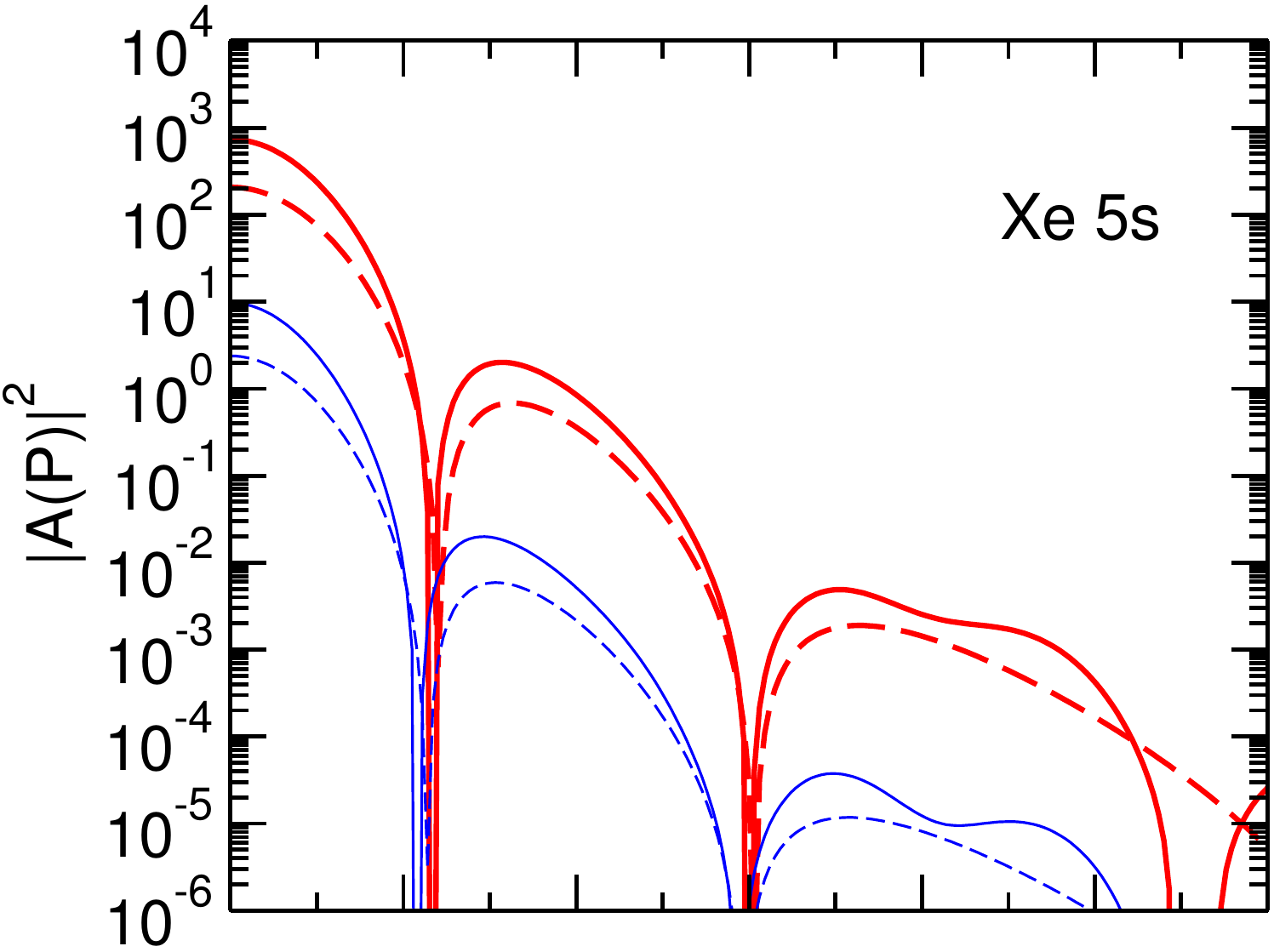}\\
\includegraphics[width=0.45\textwidth]{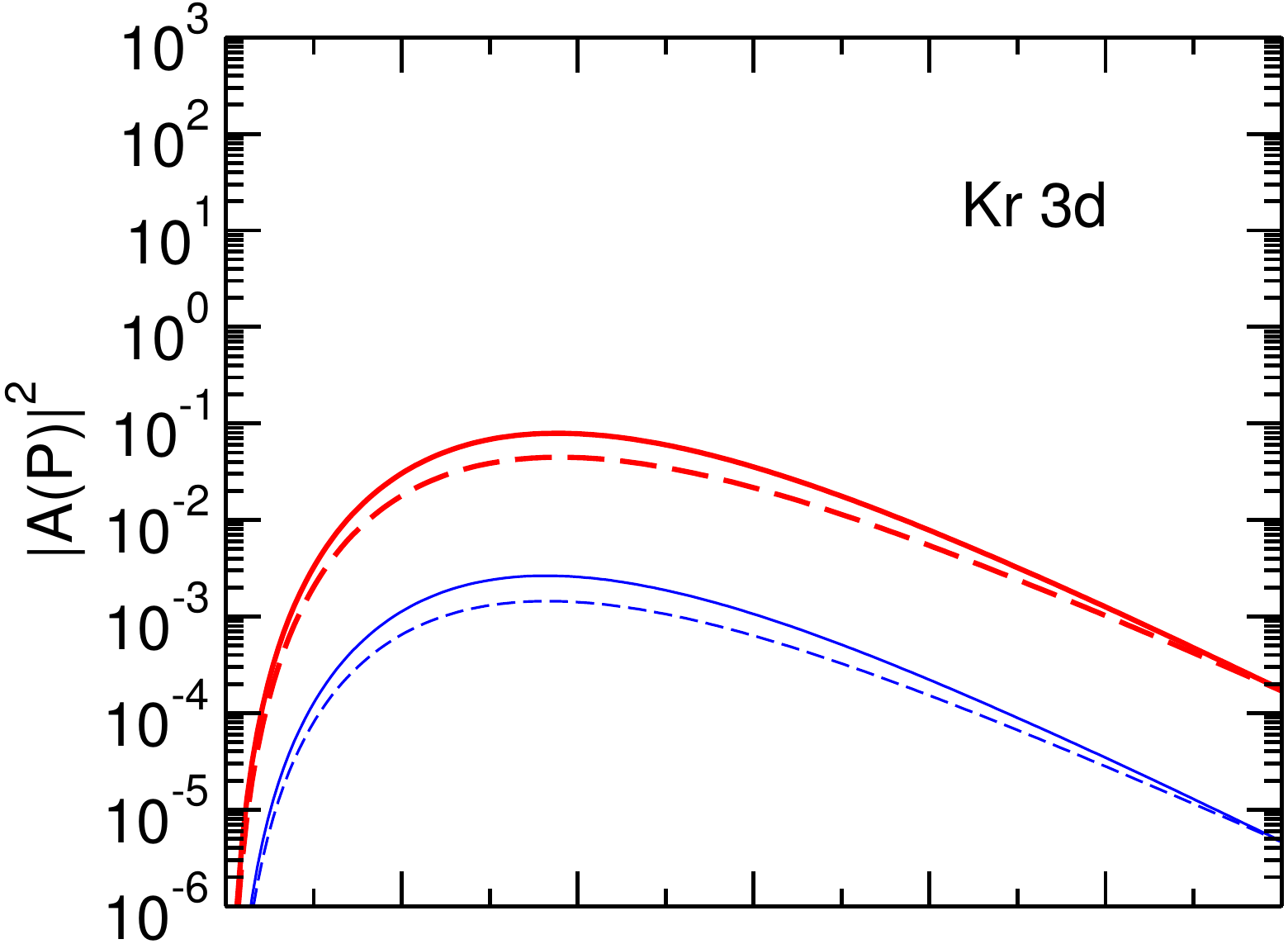}~~~~
\includegraphics[width=0.45\textwidth]{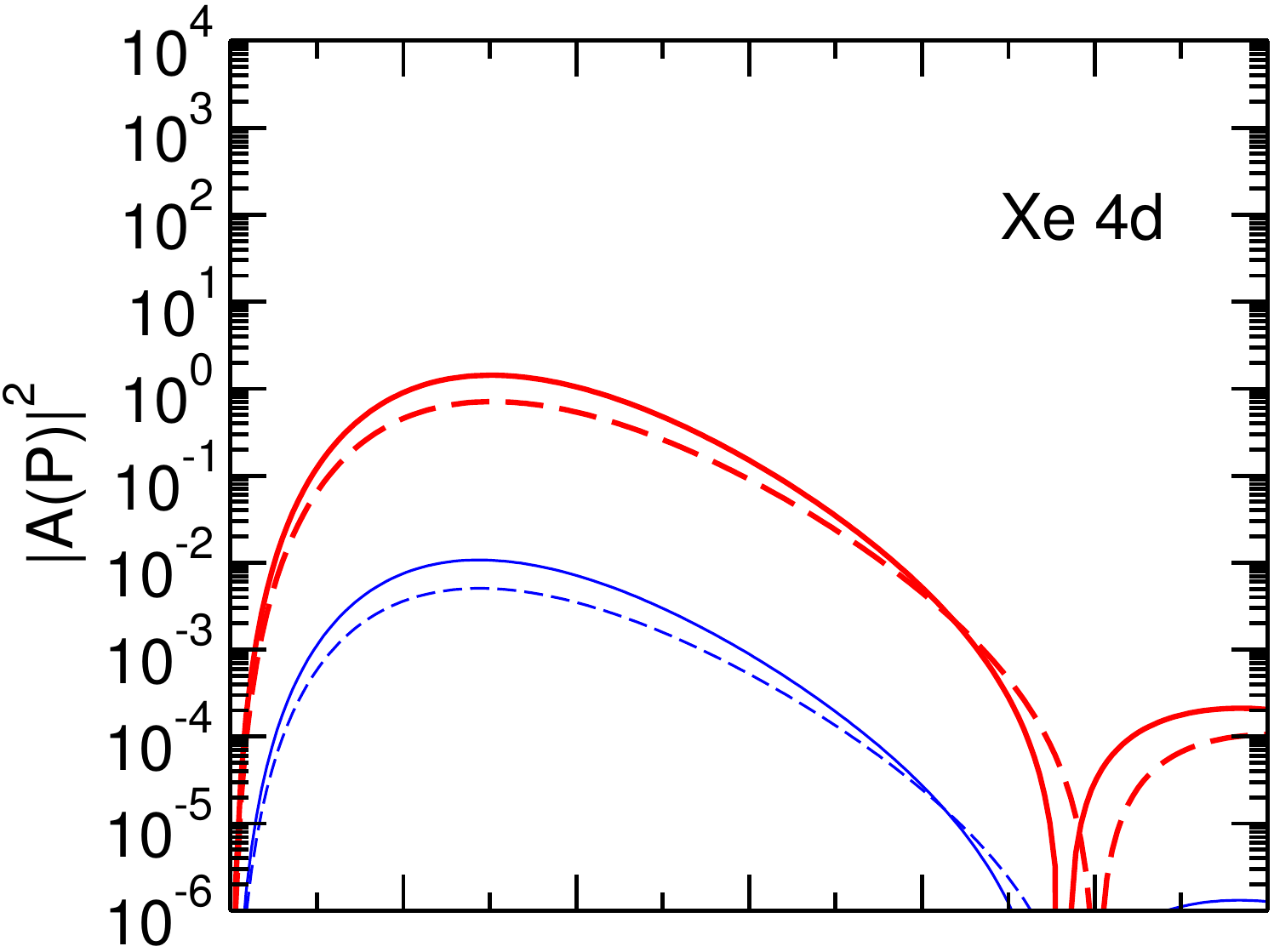}\\
\includegraphics[width=0.45\textwidth]{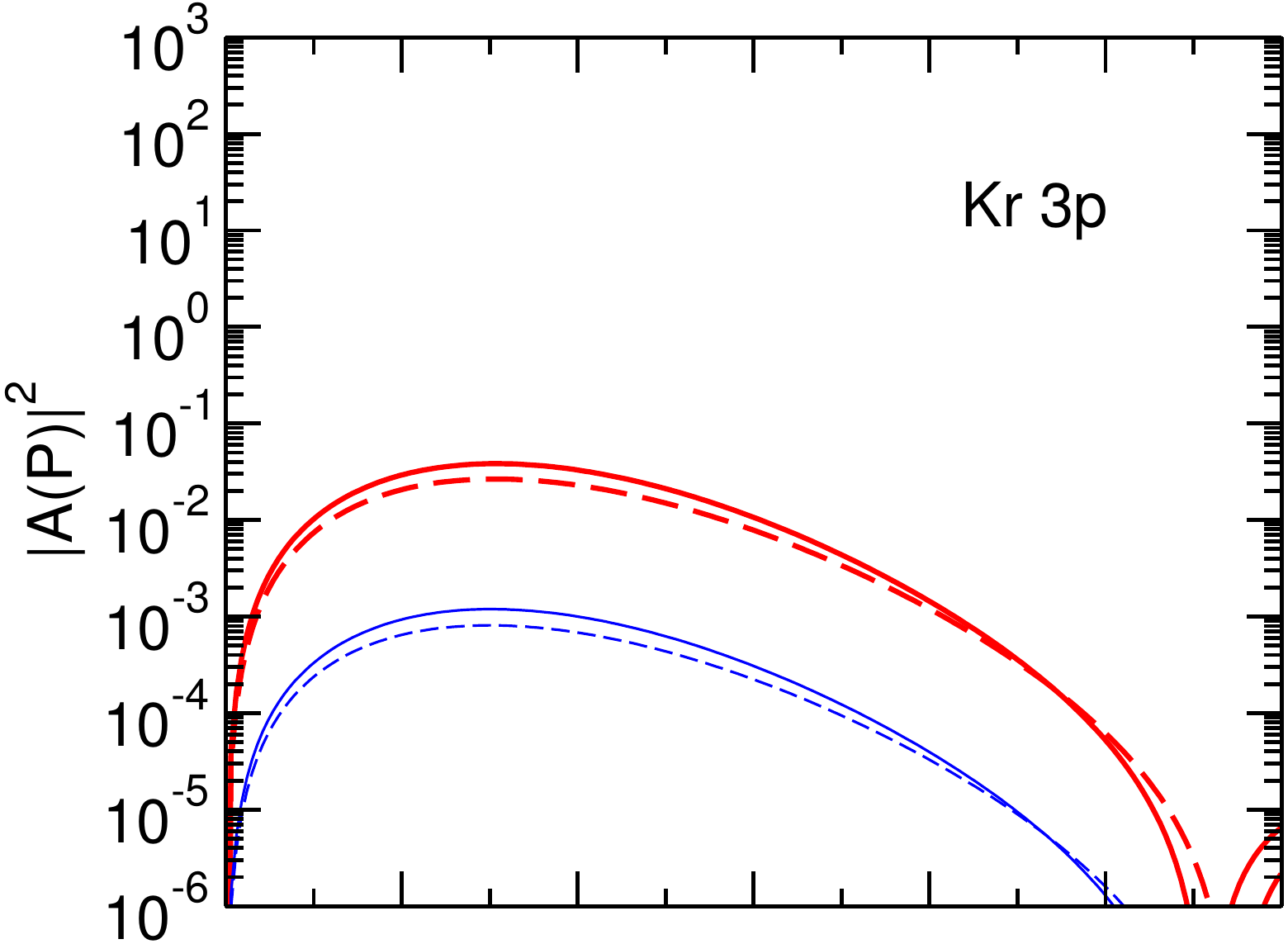}~~~~
\includegraphics[width=0.45\textwidth]{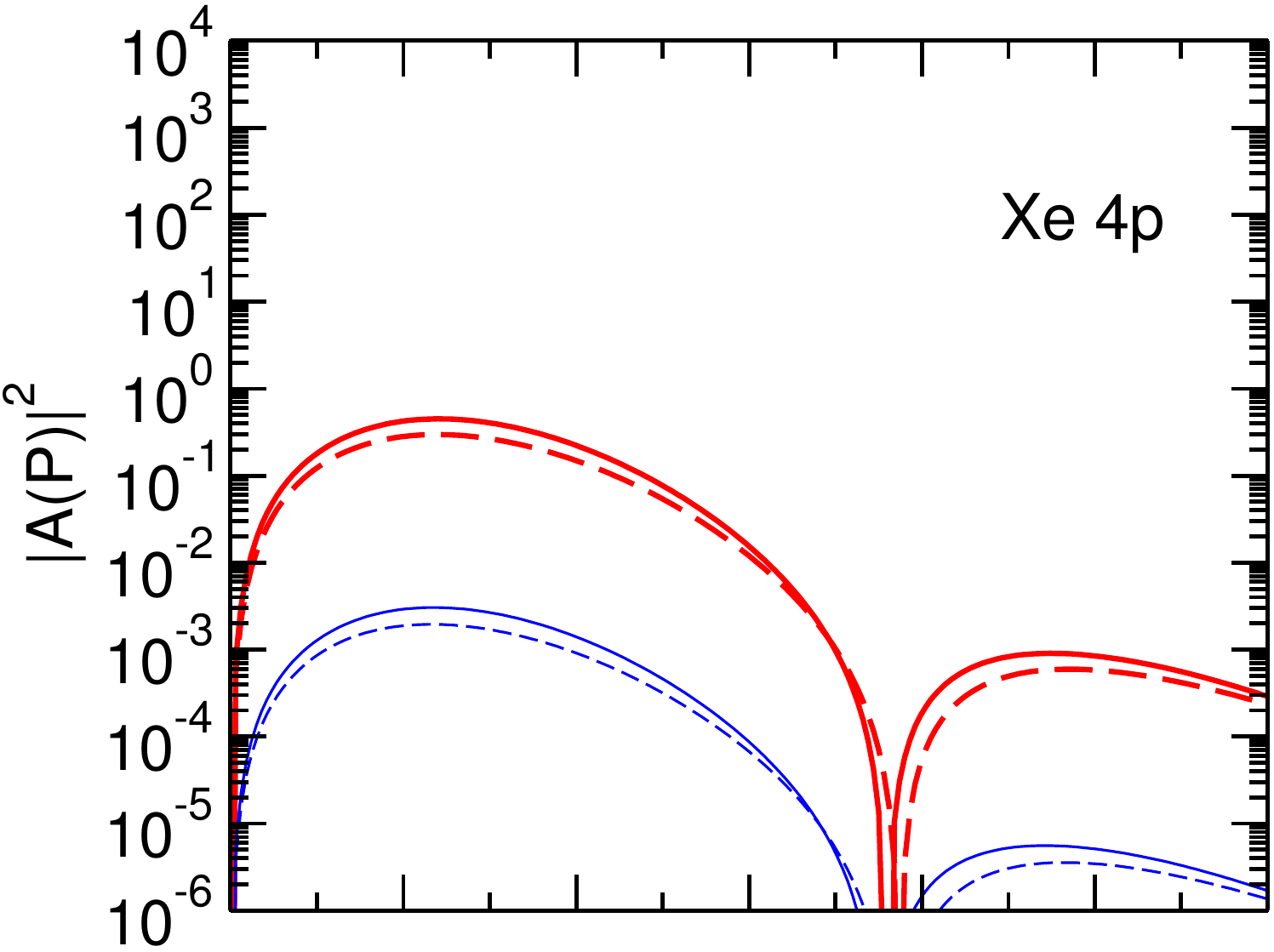}\\
\includegraphics[width=0.45\textwidth]{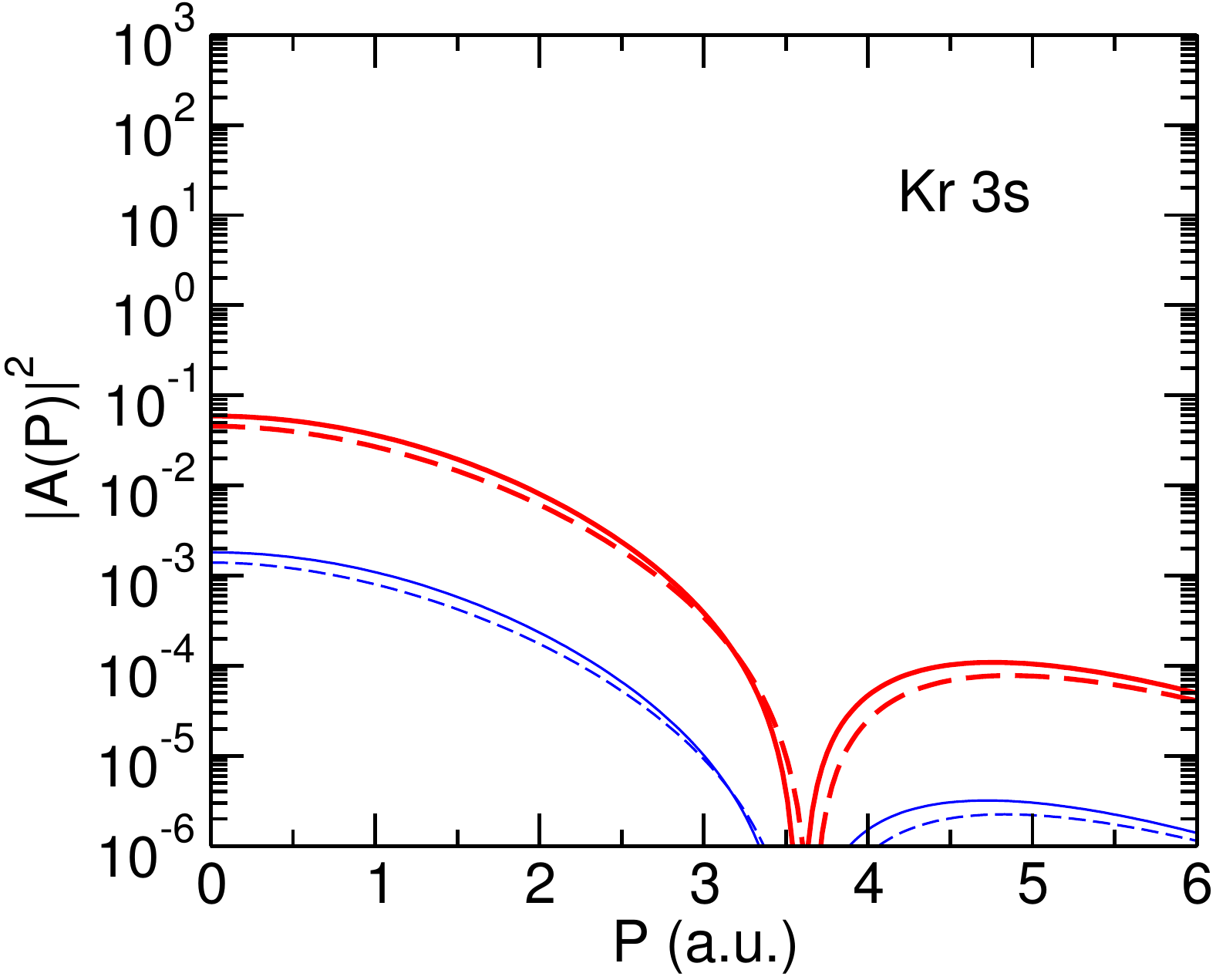}~~~~
\includegraphics[width=0.45\textwidth]{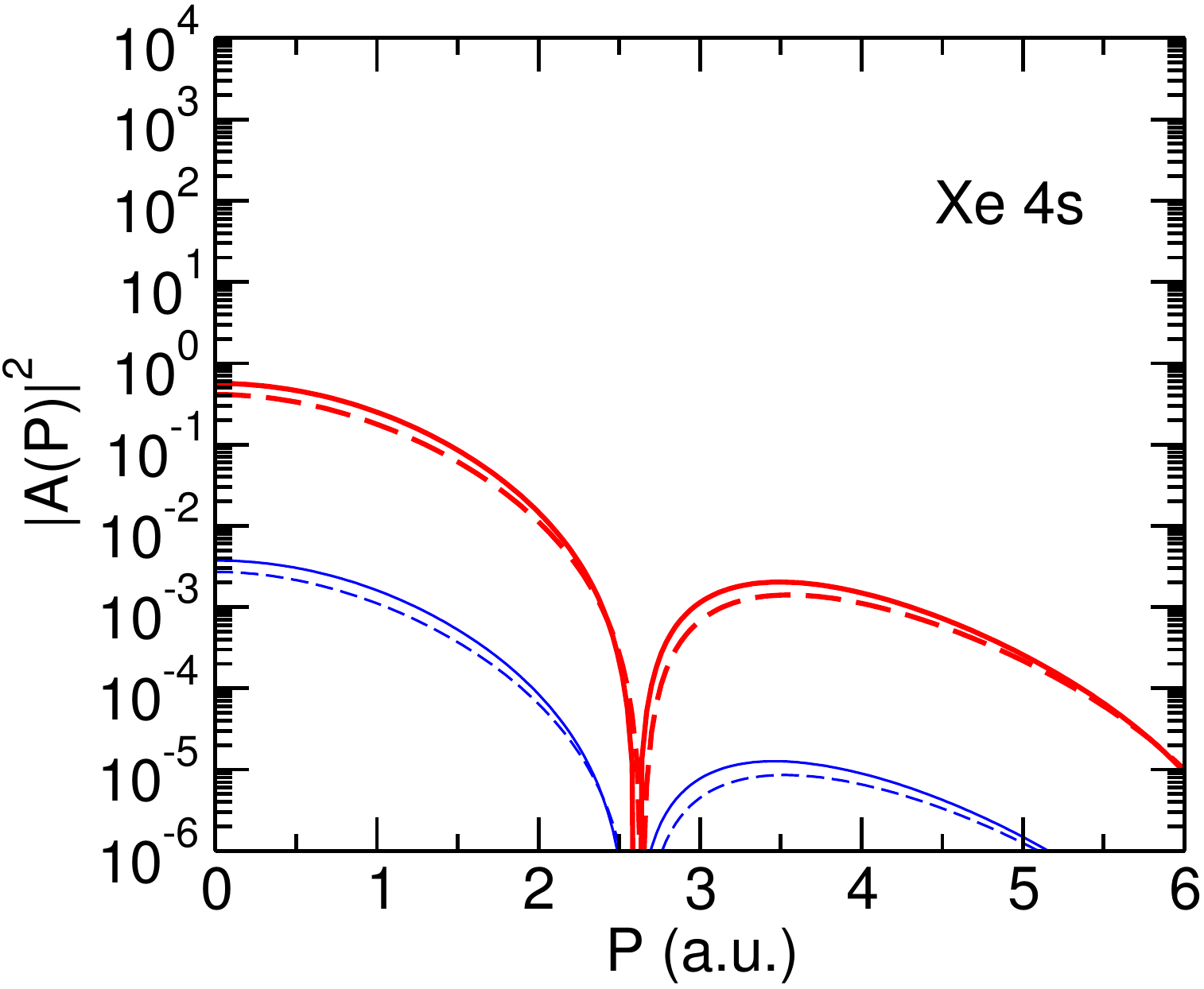}\\
\caption{Annihilation momentum densities as a function of the total $2\gamma$ momentum $P$, for s-wave positron annihilation in Kr and Xe. Various lines as described in Fig.~\ref{fig:he}
}
\label{fig:krxe_amp}       
\end{figure}

Figures \ref{fig:he} and \ref{fig:krxe_amp} show the AMD
$|A_{n\eps}({\bf P})|^2$ [spherically averaged, as in Eq.~(\ref{eqn:gammaspectra})]
for thermal ($k=0.04$ a.u.) s-wave positrons annihilating on individual core and valence subshells of the noble gase atoms, calculated using different approximations for the annihilation amplitude and positron wavefunction.
The range of two-$\gamma $ momenta $P=0$--6 a.u. corresponds to the maximum Doppler energy shift $\epsilon \approx 11$ keV. 

The simplest approximation shown uses the zeroth-order (IPA) annihilation amplitude (\ref{eq:Ank0}) with positron wavefunctions in the static field of the HF atom. Better approximations involve using the full annihilation vertex of Fig.~\ref{fig:anndiags} [Eq.~(\ref{eqn:annampgeneral})], or the best (Dyson) positron wavefunction, or both. In general, including correlations of either types increases the AMD and the annihilation probability. 

General trends are observed throughout the noble-gas sequence. The AMD are broader for the core orbitals for which the typical electron momenta are greater, leading to greater Doppler shifts. The core AMD (and the core annihilation probabilities \cite{DGG:2015:core,DGG_corelong}) also have noticeably smaller magnitudes compared with those of the valence electrons.
For all electron orbitals whose radial wavefunctions have nodes (e.g., 2s in Ne, 2s, 3s and 3p in Ar, etc.) the AMD display deep minima related to the nodes of the annihilation amplitude $A_{n\eps}({\bf P})$. Their number and positions are related to the number and positions of the nodes in the orbital's radial wavefunction (i.e., the radial nodes that occur closer to the nucleus result in the nodes of $A_{n\eps}({\bf P})$ at higher momenta). This behaviour is easy to understand from the zeroth-order amplitude (\ref{eq:Ank0}), which is the
Fourier transform of the product of the electron and positron wavefunctions. For low positron energies, its wavefunction inside the atom decreases monotonically towards the nucleus (suppressed by the repulsive electrostatic potential at smaller distances), and has no nodes. Hence, the nodal structure of the annihilation amplitude is determined by the behaviour of the electron wavefunction. Inclusion of the correlation corrections to the annnihilation vertex, as described by Eq.~(\ref{eqn:annampgeneral}), leads only to a small shift in the positions of the nodes.

For a given approximation for the annihilation vertex, the AMD calculated using the positron Dyson orbitals (red curves) are larger than those calculated using the HF positron wavefunction (blue curves). The corresponding enhancement is nearly the same for the valence and core orbitals in a given atom. It ranges from $\sim 2$ in Helium to $\sim 100$ in Xe, and is only weakly dependent on $P$. (Note that the AMD are plotted on the logarithmic scale.) This enhancement is due to the action of the attractive correlation potential $\hat \Sigma _\eps$ on the positron (Dyson) wavefunction. It leads to a build-up of the positron density in the vicinity of the atom and better overlap with the atomic electron density. This effect is stronger for the heavier, more polarizable atoms, leading to great enhancements. In Ar, Kr and Xe the attractive positron-atom potential supports low-lying virtual s levels, leading to a characteristic resonant growth of the annihilation rates at low positron energies \cite{PhysScripta.46.248,dzuba_mbt_noblegas,DGG_posnobles,PhysLett.13.300}.

When the vertex corrections are included in the annihilation amplitude (solid curves), the AMD is enhanced above the zeroth-order (IPA) result (dashed curves). This is due to the Coulomb attraction within the annihilating pair, which increases the probability of finding the electron and positron at the same point in space. The size of the enhancement is similar for the HF and Dyson positron wavefunctions. At the same time, the enhancement is much greater for the valence electrons than for the core electrons, as the former are more easily perturbed by the positron's Coulomb field. For the core electrons, the vertex correction is dominated by the first-order diagram Fig.~\ref{fig:anndiags} (b) (similar to the case of hydrogen-like ions \cite{DGG_hlike}). For the valence electrons the nonperturbative $\Gamma$-block contribution Fig.~\ref{fig:anndiags} (c) is also very important.

From Figs.~\ref{fig:he} and \ref{fig:krxe_amp} one can also see that
the vertex enhancement is significantly stronger at low momenta $P$ of the electron-positron pair (which leads to narrowing of the $\gamma$-ray 
spectra in comparison with those obtained with the zeroth-order amplitude  \cite{0953-4075-39-7-008,DGG:2015:core}). This can be seen most clearly in the AMD of the valence electrons. Their high-$P$ content is due to positron annihilation with the electron when the latter is closer to the nucleus, and where its local velocity is higher, making it less susceptible to the positron's attraction. (Note that for large $P$ the calculated $\Gamma$-block vertex corrections contain numerical errors which manifest themselves as extra oscillations visible in AMD for valence electrons. This, however, has a negligible effect on the annihilation spectra, since the AMD for the valence electrons at such momenta are very small.) The vertex enhancement is considered in more detail below.

We conclude this section by noting that calculations of the corresponding $\gamma$ spectra were reported in \cite{DGG:2015:core, DGG_corelong}. They showed excellent agreement with the measured spectra for Ar, Kr and Xe \cite{PhysRevLett.79.39}, and firmly established the fraction of core annihilation for these atoms.

\section{Vertex enhancement factors}

The enhancement of the AMD and annihilation rates due to the correlation corrections to the vertex with respect to those obtained using the zeroth-order (IPA) approximation [cf. Eqs.~(\ref{eqn:annampgeneral}) and (\ref{eq:Ank0})] can be parameterized through so-called \emph{vertex enhancement factors}. The MBT enables a direct \emph{ab initio} calculation of `exact' EF: one simply compares the results obtained with the annihilation amplitude calculated in different approximations, as described here.

The EF were introduced originally to correct the IPA annihilation rates for positrons in condensed matter (see, e.g., \cite{RevModPhys.66.841} and references therein),
\begin{eqnarray}\label{eq:ann_rate}
\lambda =\pi r_0^2c\int n_-({\bf r})n_+({\bf r})\gamma ({\bf r})d^3{\bf r},
\end{eqnarray}
where $n_-({\bf r})$ and $n_+({\bf r})$ are the electron and positron densities, respectively, and $\gamma ({\bf r})$ is the EF. The latter is typically computed for a uniform electron gas (e.g., using MBT \cite{Arponen1979343,0305-4608-9-12-009}) and parameterized in terms of the electron density, e.g., as
$\gamma =1+1.23r_s-0.0742r_s^2+\frac16 r_s^3$, where $r_s=(3/4\pi n_-)^{1/3}$ \cite{PhysRevB.51.7341} (see also \cite{PhysRevB.34.3820,PhysRevB.48.9828}). An approximation commonly used to account for the vertex enhancement of the IPA annihilation amplitude (\ref{eq:Ank0}) is
\cite{0305-4608-17-6-011,PhysRevB.54.2397}
\begin{eqnarray}\label{eqn:defenhance}
A_{n\eps}({\bf P})=\int e^{-i{\bf P}\cdot{\bf r}} \psi_{\eps}({\bf r})\psi_n({\bf r}) \sqrt{\gamma ({\bf r})}d^3{\bf r}.
\end{eqnarray}
However, this method is known to give spurious effects in the high-momentum regions of the $\gamma $ spectra \cite{PhysRevB.54.2397}.

The general MBT expression for the annihilation amplitude in a finite rather than infinite and homogeneous system, Eq.~(\ref{eqn:annampgeneral}), shows that the correlation contribution to the vertex is nonlocal, i.e., it involves the positron and electron wavefunctions $\psi_{\eps}({\bf r}_1)$ and $\psi_n({\bf r}_2)$ at different points in space. The corresponding enhancement is described by the three-point function
$\tilde\Delta _\eps ({\bf r};{\bf r}_1,{\bf r}_2)$. This allows one to formally define the EF for the electron in orbital $n$ and positron of energy $\eps $ by
\begin{eqnarray}\label{eqn:bargammapformal}
\sqrt{\gamma_{n\eps}({\bf r})}\equiv 1+\frac{\iint \tilde\Delta _\eps ({\bf r};{\bf r}_1{\bf r}_2)\psi_{\eps}({\bf r}_1)\varphi_n({\bf r}_2) d^3{\bf r}_1d^3{\bf r}_2}{\psi_{\eps}({\bf r})\varphi_n({\bf r})}.
\end{eqnarray}
However, the presence of nodes in the wavefunctions in the denominator renders this quantity of limited use and we must opt for a more pragmatic approach.

It is clear from Figs.~\ref{fig:he} and \ref{fig:krxe_amp} that the vertex enhancement of the AMD $|A_{n\eps}({\bf P})|^2$, i.e.,
full-vertex results compared with zeroth-order, has a weak dependence on the momentum $P$ (except near the nodes of the amplitude). This momentum dependence of the vertex enhancement has little effect on the annihilation $\gamma$ spectra for the noble-gas atoms, especially for the core orbitals \cite{DGG_corelong}.
It is thus instructive to define a two-$\gamma$ momentum-averaged vertex EF as the ratio of the full-vertex partial annihilation rate to that calculated using the zeroth-order (IPA) vertex:
\begin{eqnarray}\label{eqn:ef}
\bar\gamma_{nl}(k)=\frac{Z^{(0+1+\Gamma)}_{{\rm eff}, nl}(k)}{Z^{(0)}_{{\rm eff}, nl}(k)},
\end{eqnarray}
where the superscript denotes the vertex order (see Fig.~\ref{fig:anndiags}) and $nl$ labels the subshell of the electron that the positron annihilates with.
Similar EF are commonly used to analyse and predict the annihilation rates and $\gamma$ spectra in solids (see, e.g., \cite{JPhysCM.1.10595,PhysRevB.51.4176,Barbiellini1997283,PhysRevB.56.7136,PhysRevB.54.2397,PhysRevB.73.035103,PhysRevB.58.10475}).
The true spectrum for annihilation on a given subshell for a given positron momentum can then be approximated by
\begin{eqnarray}\label{eqn:spectra_reconstructed}
w_{nl}(\epsilon)\approx \bar\gamma_{nl} w^{(0)}_{nl}(\epsilon)
\end{eqnarray}
where $w^{(0)}_{nl}$ is the $\gamma$ spectrum calculated using the zeroth-order vertex. Accurate reconstruction of the true spectra for s-wave thermal positrons using Eqn.~(\ref{eqn:spectra_reconstructed}) has been demonstrated for noble-gas atoms in \cite{DGG_corelong}.

In a recent paper \cite{DGG:2015:core} we showed that for thermal s-wave positrons ($k=0.04$~a.u.) $\bar\gamma_{nl}$ follows a near-universal scaling with the orbital ionization energy $I_{nl}$,
\begin{eqnarray}\label{eqn:uscale}
\bar\gamma_{nl} = 1+\sqrt{A/I_{nl}}+ (B/I_{nl})^{\beta},
\end{eqnarray}
where $A$, $B$ and $\beta$ are constants \footnote{For HF positron wavefunctions the values of the parameters are $A=1.54~\text{a.u.}=43.6~\text{eV}$, $B=0.92~\text{a.u.}=20.5~\text{eV}$, and $\beta=2.54$.
For Dyson positron wavefunctions the values are $A=1.31~\text{a.u.}=36~\text{eV}$, $B=0.83~\text{a.u.}=27.0~\text{eV}$, and $\beta=2.15$ \cite{DGG:2015:core}.}.
The second term on the RHS of Eqn.~(\ref{eqn:uscale}) describes the effect of the first-order correction, Fig.~\ref{fig:anndiags} (b). Its scaling with $I_{nl}$ was motivated by the $1/Z$ scaling for positron annihilation in hydrogenlike ions \cite{DGG_hlike}. The third term is phenomenological and describes the effect of the $\Gamma$-block correction that is particularly important for the valence subshells.

Here we extend the calculations of the enhancement factors to s-, p- and d-wave positrons with momenta up to the positronium-formation threshold. At small, e.g., room-temperature, thermal positron momenta $k\sim 0.04$ a.u., the contributions of the positron p and d waves to the annihilation rates are very small, owing to $Z_{\rm eff}(k)\propto k^{2\ell}$ low-energy behaviour. (This is a manifestation of the suppression of the positron wavefunction in the vicinity of the atom by the centrifugal potential $\ell (\ell +1)/2r^2$.) However, for higher momenta close to the Ps-formation threshold, the s-, p- and d-wave contributions to the annihilation rates become of comparable magnitude
(see Fig. 16 and Tables III--VII in Ref.~\cite{DGG_posnobles}).

\begin{figure}[ht!]
~~~~\includegraphics[width=1\textwidth]{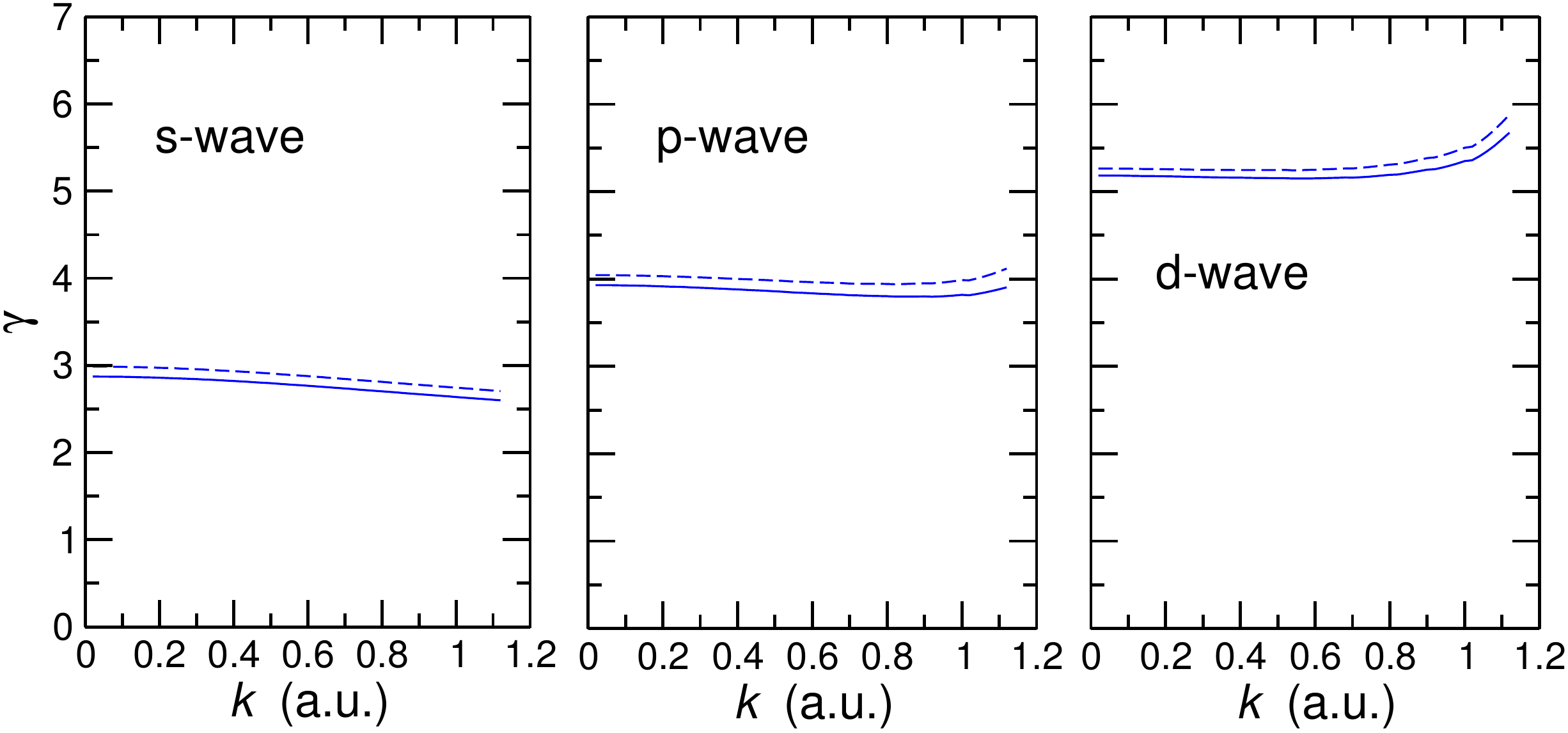}
\caption{Enhancement factors for s-, p- and d-wave positrons annihilating on the 1s electrons in He, obtained with HF (dashed lines) and Dyson (solid lines) positron wavefunctions.
}
\label{fig:he_gam}       
\end{figure}

\begin{figure}[ht!]
\includegraphics[width=1\textwidth]{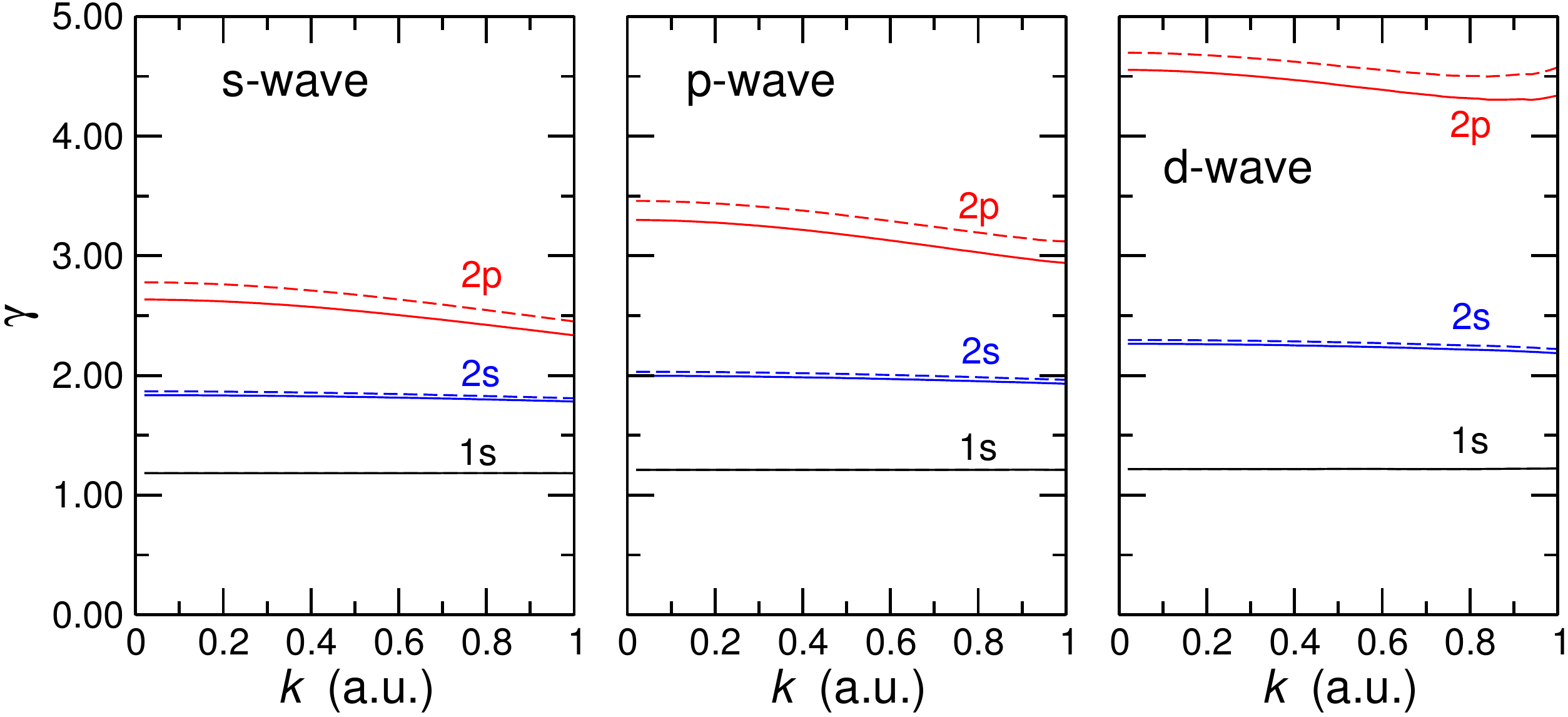}
\caption{Enhancement factors for s-, p- and d-wave positrons annihilating on the 1s, 2s and 2p subshells in Ne, obtained with  HF (dashed lines) and Dyson (solid lines) positron wavefunctions.
}
\label{fig:ne_gam}       
\end{figure}

\begin{figure}[ht!]
\includegraphics[width=1\textwidth]{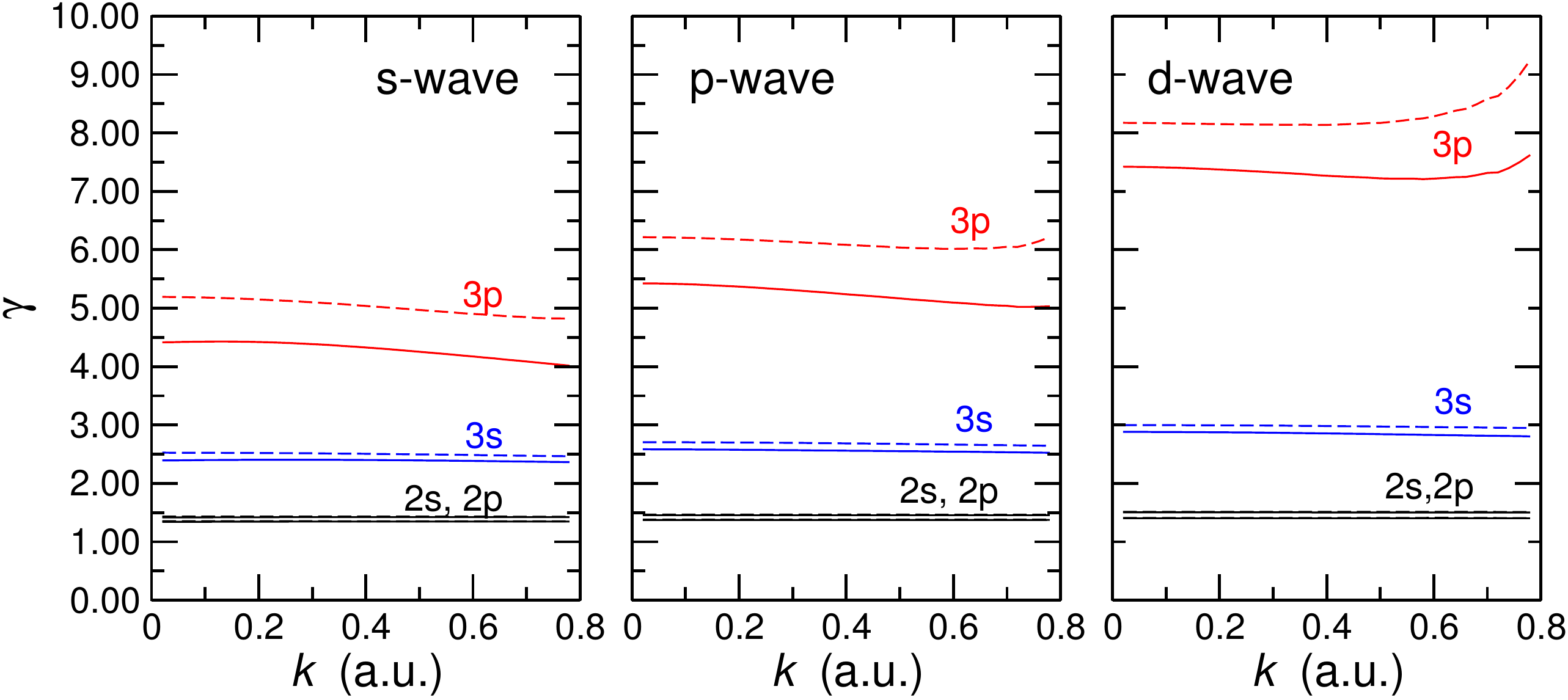}
\caption{Enhancement factors for s-, p- and d-wave positrons annihilating on the 2s, 2p, 3s and 3p subshells in Ar, obtained with  HF (dashed lines) and Dyson (solid lines) positron wavefunctions.
}
\label{fig:ar_gam}       
\end{figure}

\begin{figure}[ht!]
\includegraphics[width=1\textwidth]{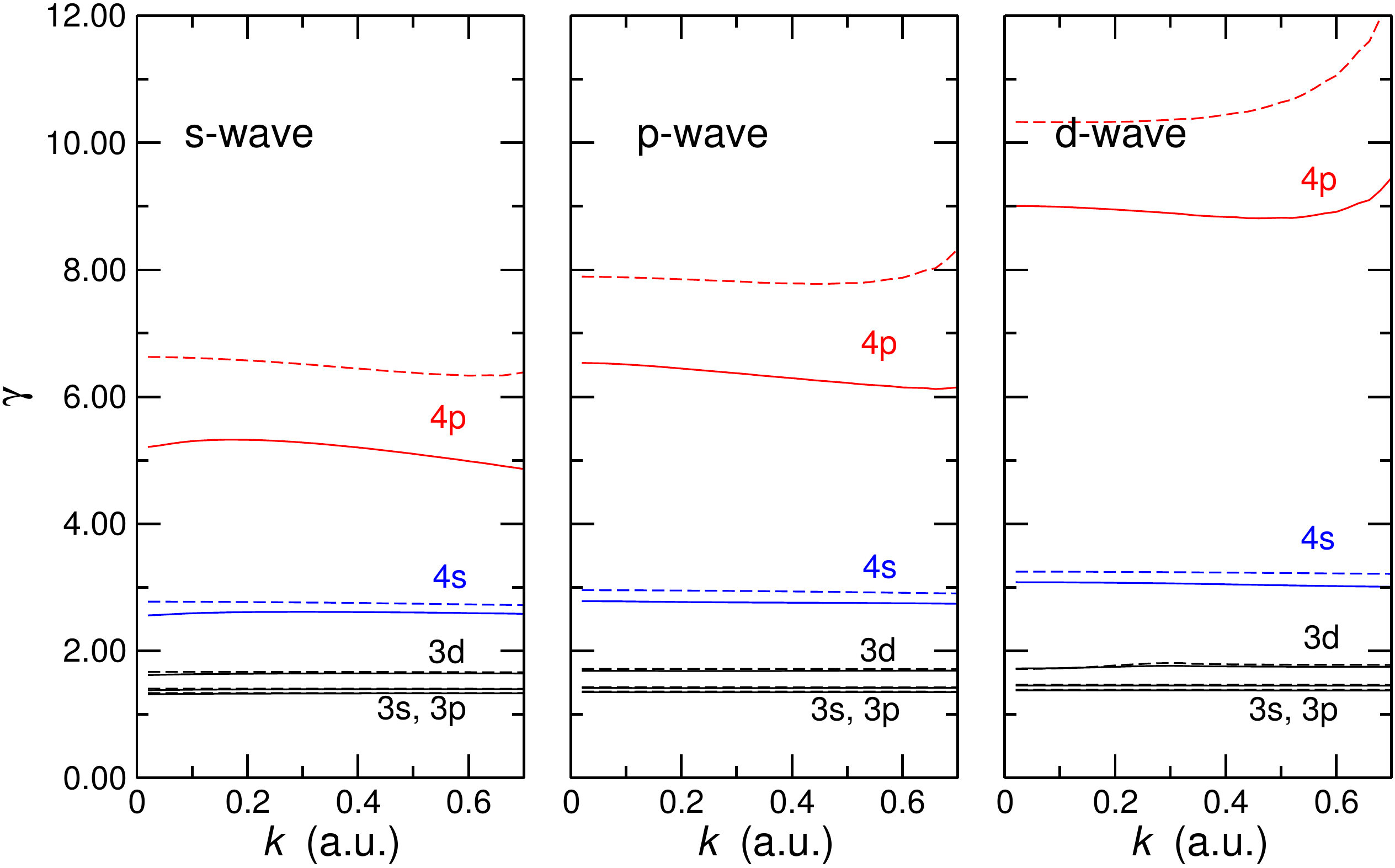}\\[2ex]
\caption{Enhancement factors for s-, p- and d-wave positrons annihilating on the 3s, 3p, 3d, 4s and 4p subshells in Kr, obtained with  HF (dashed lines) and Dyson (solid lines) positron wavefunctions.}
\label{fig:kr_gam}       
\end{figure}

\begin{figure}[ht!]
\includegraphics[width=1\textwidth]{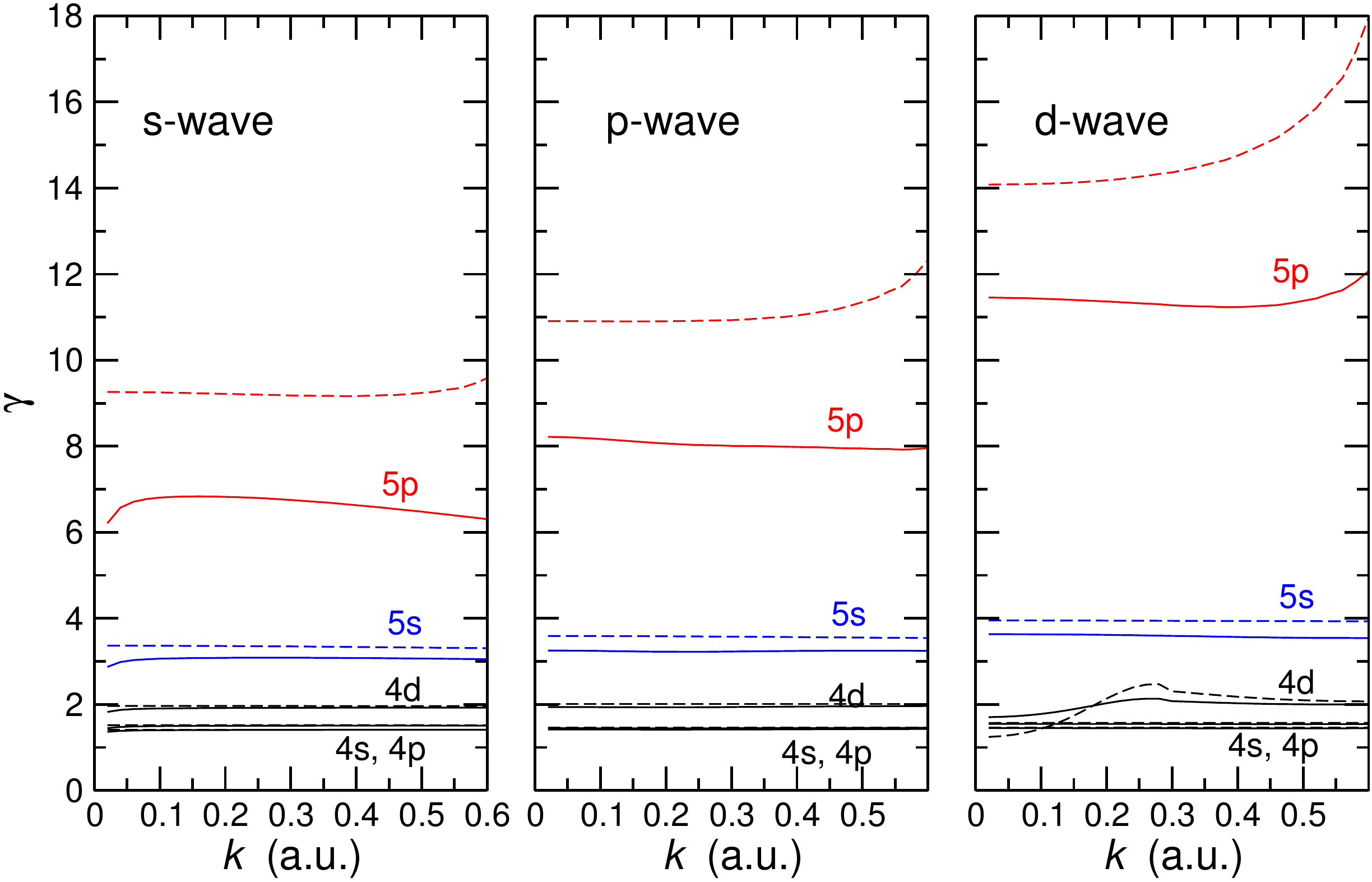}\\[2ex]
\caption{Enhancement factors for s-, p- and d-wave positrons annihilating on the 4s, 4p, 4d, 5s and 5p subshells in Xe, obtained with  HF (dashed lines) and Dyson (solid lines) positron wavefunctions.
}
\label{fig:xe_gam}       
\end{figure}

Figures~\ref{fig:he_gam}--\ref{fig:xe_gam} show the enhancement factors for positron annihilation with electrons in the valence ($n$p and $n$s) and core [$(n-1)$s, $(n-1)$p, $(n-1)$d, as applicable] orbitals of the He, Ne, Ar, Kr, and Xe, as functions of the positron momentum.

The EF for positron annihilation with 1s electrons in He 
(Fig.~\ref{fig:he_gam}) are 2.6--3.0 for the s-wave, 3.8--4.1 for the p-wave, and 5.2--5.9 for the d-wave. They show only a weak dependence on the positron momentum, which is a typical feature of all the data. There is also little difference between the EF obtained with the static-field (HF) positron wavefunctions (dashed lines) and those found using the positron Dyson orbitals (solid lines). This is in spite of the fact that the use of the correlated Dyson positron wavefunctions increases the AMD (and the annihilation rates \cite{DGG_posnobles}) by almost an order of magnitude for $s$-wave positrons (Fig.~\ref{fig:he}).

The weak dependence of the EF on the positron energy and the type of positron wavefunction used is related to the nature of the vertex enhancement. The intermediate electron and positron states in diagrams Fig.~\ref{fig:anndiags} (b) and (c) that describe the short-range vertex enhancement ($\nu$, $\mu$, $\nu_1$, $\mu_1$, etc.) are highly virtual, i.e., have relatively large energies. For example, the energy denominator of diagram Fig.~\ref{fig:anndiags} (b) is $\eps -\eps _\nu -\eps _\mu +\eps _n$ (see Refs.~\cite{0953-4075-39-7-008,DGG_corelong}). Estimating the typical electron and positron energies as $\eps _{\nu,\mu} \sim |\eps _n|$ (the ionization energy of electron orbital $n$), we see that for few-electronvolt positrons, the positron energy $\eps $ can be neglected. For the same reason, the vertex correction function $\tilde \Delta _\eps ({\bf r};{\bf r}_1,{\bf r}_2)$ is only weakly nonlocal, i.e., it is large only for $|{\bf r}_1-{\bf r}_2|\ll |{\bf r}_{1,2}|\sim |{\bf r}|$ (see the ``annihilation maps'' in Figs. 4.14--4.16 of Ref.~\cite{Ludlow_thesis}). The situation becomes different at large momenta close to the Ps formation threshold. Here the p- and d-wave EF show an upturn related to the virtual Ps formation becoming ``more real'' \cite{PhysRevLett.88.163202}. This is also seen in $\bar \gamma_\text{np}$ for heavier atoms.

The increase of the EF with the positron orbital angular momentum $\ell$ seen in Fig.~\ref{fig:he_gam} can be related to the behaviour of the low-energy positron wavefunctions near the atom. Due to the action of the centrifugal potential, the p- and d-wave radial wavefunctions are suppressed as $(kr)^{\ell }$ with $\ell =1$ and 2, compared with the s wave.
The nonlocal correlation corrections Fig.~\ref{fig:anndiags} (b) and (c) ``help'' the positron to pull the atomic electron towards larger distances, which has a greater advantage for the higher partial waves. 

It is interesting to compare the values of $\bar \gamma _{1s}$ for He with the EF for positron annihilation with atomic hydrogen: 6--7, 10--12, and 15--17, for the s-, p-, and d-wave positrons, respectively, with $k\leq 0.4$~a.u. (see Fig.~13 in Ref.~\cite{PhysRevA.70.032720}). The greater values of the EF for hydrogen are related to the smaller binding energy of the 1s electron in hydrogen (13.6~eV) compared with that in He (24.6~eV). The vertex corrections are generally greater for the more weakly bound electron, as it is more easily perturbed by the positron's Coulomb interaction. The same trend will be seen throughout the noble-gas-atom sequence, with more strongly bound electron orbitals, in particular those in the core, displaying smaller EF [cf. Eq.~(\ref{eqn:uscale})].

Turning to Ne (Fig.~\ref{fig:ne_gam}), we observe that the EF for the outer valence 2p subshell are slightly smaller than those for 1s in He,
in spite of the binding energy of the 2p electrons (21.6~eV) being lower than that of He 1s. Ne also has the broadest $\gamma $ ray spectrum of all the noble gases (see AMD in Fig.~\ref{fig:he}, and the data for the calculated and measured spectra \cite{DGG_corelong,PhysRevA.55.3586}). The latter indicates that the 2p electrons in Ne have large typical momenta, which makes the correlation correction to the annihilation vertex relatively small. The EF for the inner valence 2s subshell is around 2, while for the deeply bound 1s electrons $\bar \gamma _\text{1s}\approx 1.2$. We also note that for the core orbitals the values of the EF for the positron s, p and d waves are quite close.
This is in fact a general trend observed for all atoms that the \emph{relative difference} between the values of $\bar \gamma _{nl} -1$ for the positron s, p and d waves is becoming small with the increase in the binding energy.
The smaller effect of the orbital angular momentum of the positron on the EF for  core orbitals is due to the vertex correction becoming ``more local'', and hence, less sensitive to the variation of the positron radial wavefunction.

The EF in Ar, Kr and Xe (Figs.~\ref{fig:ar_gam}, \ref{fig:kr_gam} and \ref{fig:xe_gam}) become progressively larger, for both the valence and core electrons. For example, the vertex EF for s-wave positron annihilation with the outer valence $n$p electrons increases from $\bar \gamma _\text{3p}=5.2$ (Ar), to
$\bar \gamma _\text{4p}=6.6$ (Kr), to $\bar \gamma _\text{5p}=9.2$ (Xe) (for the HF positron wavefunction at low momenta $k\lesssim 0.1$~a.u.). The EF for the $(n-1)l$ core orbitals also increase to $\bar \gamma _{(n-1)l}\sim 1.5$--2, with the values for the 3d and 4d orbitals being noticeably larger than those of the 3s/3p and 4s/4p orbitals, for Kr and Xe, respectively.

Another feature of the data is the growing difference between the EF for the np electrons obtained with the Dyson positron wavefunction (solid lines) and those found using the static-field (HF) positron wavefunction (dashed lines). This effect is related to the increase in the strength of the positron-atom correlation potential $\hat \Sigma _\eps$ for the heavier noble-gas atoms \cite{PhysScripta.46.248,dzuba_mbt_noblegas,DGG_posnobles}. For s-wave positrons it results in the creation of positron-atom virtual levels \cite{quantummechanics} whose energies
$\eps =\kappa ^2/2$ become lower for heavier atoms, with values of $\kappa =-0.23$, $-0.10$ and $-0.012$~a.u. for Ar, Kr and Xe, respectively \cite{DGG_posnobles}. This is accompanied by a rapid growth of the positron wavefunction near the atom, with the Dyson orbitals being enhanced by a factor $\sim 1/\kappa $ compared to the static-field positron wavefunctions at low energies. Hence, the inclusion of the correlation potential makes the radial dependence of the positron wavefunction more vigorous. This is evidenced by some broadening of the $\gamma $ spectra obtained with the Dyson rather than the HF positron wavefunction \cite{DGG_corelong}. This also results in a reduction of the EF, which is most noticeable for the valence electrons, and is largest in Xe, which has the strongest correlation potential for the positron.

Besides the rise in the valence EF for high positron momenta (related to the proximity of the Ps-formation threshold), one other exception from the weak momentum-dependence of the EF is seen at low momenta for d-wave positron annihilating on the 3d and 4d orbitals in Kr and Xe. The enhancement factors in this case are approximately constant from the Ps-formation threshold down to $\sim 0.3$ a.u., but then deviate at lower momenta, especially in Xe. Both the zeroth-order and full-vertex $Z_{\rm eff}$ values for these orbitals are calculated to be smooth functions of $k$. 
However, they obey the $\sim k^4$ behaviour and become very small at low $k$ (e.g., for Xe, using the Dyson wavefunction we find $Z_{\rm eff,4d}\sim 10^{-3}$ at $k\sim 0.3$ a.u., decreasing to $\sim 10^{-7}$ for $k\sim 0.03$ a.u. It appears that for such small $k$ numerical inaccuracies arise in the calculation of the $\Gamma $-block contribution, leading to errors when evaluating the ratio in Eq. (13).

\section{Conclusions}

We used many-body theory methods to calculate the annihilation momentum densities and vertex enhancement factors for s-, p- and d-wave positrons annihilating on valence and core electrons in noble-gas atoms. The general trends of the EF is their weak dependence on the positron momentum and decrease with the increasing binding energy. We find that the type of the positron wavefunction used, i.e., Dyson orbital which accounts for the positron-atom correlation attraction vs repulsive, static-field (HF) wavefunction, has relatively little effect on the EF, except for the valence orbitals in most polarisable targets. We also find a relatively weak dependence of the EF for the core and inner-valence electrons on the positron's orbital angular momentum. 

The weak momentum-dependence of the EF obtained in positron-atom calculations suggests that they can be used to improve the calculations of positron annihilation in more complex environments. One such system is positronium colliding with noble-gas atoms, where calculations of Ps-atom pick-off annihilation rates that neglect the short-range vertex enhancement strongly underestimate the measured rates \cite{PhysRevA.65.012509}. Another context where similar EF can be used is positron annihilation in molecules. Here there is a sharp contrast between the large amount of experimental information, including $\gamma $-spectra, for a wide range of molecule \cite{PhysRevA.55.3586} and paucity of credible theoretical data \cite{RevModPhys.82.2557,DGG_molgamma}. The positron-molecule problem is particularly interesting because the $Z_\text{eff}$ values for most polyatomic molecules show orders-of-magnitude enhancement due to resonant positron annihilation \cite{RevModPhys.82.2557}. In such molecules the positron annihilates from a temporarily formed weakly-bound state. Attempts to calculate such states using standard quantum-chemistry methods have been numerous but not very successful \cite{RevModPhys.82.2557} (i.e., there is only a small number of systems where theory and experiment can be compared, and the agreement is mostly qualitative \cite{Tachikawa_2014}).

\begin{acknowledgement}
DGG is supported by a United Kingdom Engineering and Physical Sciences Research Council Fellowship, grant number EP/N007948/1.
\end{acknowledgement}

%

\end{document}